\documentclass[12pt]{article}

\usepackage{epsfig}
\usepackage{graphicx}

\title{A theory of quantum gravity based on quantum computation}

\author
{Seth Lloyd\\
\\
\normalsize{Department of Mechanical Engineering}\\
\normalsize{MIT 3-160, Cambridge MA 02139 USA}\\
\normalsize{The Santa Fe Institute}\\
\normalsize{1399 Hyde Park Road, Santa Fe, NM 87501 USA}\\
\\
\normalsize{slloyd@mit.edu}
}

\date{}

\begin{document}


\baselineskip24pt


\maketitle

\begin{abstract}
\noindent This paper proposes a method of unifying 
quantum mechanics and gravity based on quantum computation.
In this theory, fundamental processes are described in terms of pairwise
interactions between quantum degrees of freedom.  
The geometry of space-time is a construct, derived from the underlying 
quantum information processing.  The computation
gives rise to a superposition of four-dimensional spacetimes, each of
which obeys the Einstein-Regge equations.  The theory makes
explicit predictions for the back-reaction of the metric to
computational `matter,' black-hole evaporation, holography, and quantum
cosmology.
\end{abstract}

\vskip .5in
                                                                                
Quantum computation can be thought of as a universal theory for
discrete quantum mechanics.  Quantum computers are discrete systems 
that evolve by local interactions [1], and every discrete quantum system
that evolves by local interactions, including lattice gauge theories, can
be simulated efficiently on a quantum computer [2-6]
The quantization of gravity remains one of the primary challenges
to physics [7-31].  If, at bottom, quantum gravity is a discrete, 
local quantum theory, then quantum gravity, too, should be describable 
as a quantum computation.  

Unlike conventional approaches to quantum gravity such as 
string theory [14], canonical quantization [7],
loop quantum gravity [15-20], and Euclidean quantum gravity [10] 
the theory proposed here does not set out to quantize gravity directly.  
Gravity is a theory based on geometry and distance: the normal
approach to gravity is to quantize the metric of spacetime.  In the theory
investigated here, the concept of distance is not a fundamental one. 
Instead, distances are quantities that are derived from the 
underlying dynamics of quantum systems.  For example, a
lattice gauge theory can be written as a sum of local Hamiltonians
that give interactions between the fields at different points on
the lattice.  In our theory, rather than regarding the distances
between points on the lattice as independent variables, we {\it derive} 
those distances from the quantum behavior of the underlying fields.  
As will be shown below, those derived distances automatically conform
to the laws of general relativity: that is, a lattice theory of the 
standard model can give rise directly to a theory of quantum gravity, 
without ever explicitly quantizing the gravitational field.
Quantum fluctuations in the interactions between fields translate
directly into quantum fluctuations in those distances, and consequently 
in the metric of spacetime.  Because distances are derived from
dynamics, without reference to an underlying spacetime manifold,
the resulting theory is intrinsically covariant and
background independent: the observable content of the theory
resides in the underlying computation, and is independent of 
any background or choice of coordinates.

Because of their ability to reproduce the dynamics of discrete
quantum systems, including local Hamiltonian systems such as
lattice gauge theories, quantum computers will be used here
as the fundamental system from which to derive quantum gravity.
Phrased in terms of quantum computation, what is quantized
is not the metric of spacetime; rather,
what is quantized here is information.  All observable
aspects of the universe, including the metric structure of spacetime
and the behavior of quantum fields, are derived from and arise out of an
underlying quantum computation.  The form that quantum fluctuations
in geometry take can be calculated directly from the 
quantum computation.  To paraphrase Wheeler,
`it from qubit' [12-13].    The approach of deriving geometry
from the behavior of quantum mechanical matter
is reminiscent the work of Sakharov [21]; however, as will be seen,
the method presented here for deriving gravity from the underlying quantum
dynamics differs from Sakharov's, and resolves issues such as the 
back reaction and the coupling of the metric to quantum fluctuations 
that Sakharov's theory failed to resolve.  Apart from Sakharov,
the closest existing approach to the one
taken here is that of causal sets [22-26].  
(See also [27-28]; for work relating quantum computation to loop quantum 
gravity see [13,29-30]).  

This paper is organized as follows:
\smallskip
Section (1) reviews the theory of quantum computation, and shows 
how a quantum computation can be written as a superposition of 
computational `histories,' each one of which possesses a definite 
causal structure and a definite local action.  Quantum fluctuations
in causal structure and in the local action and energy are determined
by the local dynamics of the computation. 
\smallskip
Section (2) shows that each of these histories, in turn, gives rise
to a classical discretized spacetime geometry, whose metric
is {\it deduced} from the causal structure and local action.    
The superposition of the histories in the quantum computation gives
rise to a quantum superposition of spacetimes, whose fluctuations
in causal structure and curvature are determined by the same
local computational dynamics.  
\smallskip
Section (3) applies the results of sections (1-2) to analyze 
the quantum behavior of singularities, black hole evaporation, 
the quantum back reaction, the holographic principle, 
and quantum cosmology.  
These results can be summarized as follows:

\smallskip\noindent
(a) Singularities obey a form of the cosmic censorship hypothesis.

\smallskip\noindent
(b) Black hole evaporation is either unitary or approximately
unitary: all or most information escapes from black holes
as they evaporate.  

\smallskip\noindent
(c) Because distances are determined from
underlying quantum dynamics, fluctuations in distance track
quantum fluctuations in the computational `matter': 
a quantum fluctuation in the device that determines distance
is indistinguishable from a quantum fluctuation in the distance
itself.  

\smallskip\noindent
(d) The theory proposed here is consistent with holography
and supplies a complementary principle, the quantum geometric
limit, which limits the number of elementary events, or `ops,' that can
occur within a four-volume of spacetime.  

\smallskip\noindent
(e) Finally, simple quantum
cosmologies can give rise to primordial, Planck-scale inflation,
followed by epochs of radiation dominance and matter dominance;
parts of the universe can then begin to re-inflate a reduced
rate determined by the Hubble parameter at that later epoch.
The mechanisms for primordial inflation and late-epoch inflation
are essentially the same: only the dynamically determined fundamental
length scale changes.

\section{Quantum computation}
                                                                                
Quantum computers are devices that process information in a way
that preserves quantum coherence.  The quantum information that
quantum computers process is registered on quantum degrees of
freedom, typically a `qubit' with two distinct states such as electron
spin or photon polarization.   

\subsection{The computational graph}
                                                                                
Each quantum computation corresponds to a directed, acyclic graph
$G$, the `wiring diagram' for the computation (figure 1).  
The initial vertices of the graph correspond to input states.
The directed edges of the graph correspond to quantum wires that move 
quantum information from place to place.   
The internal vertices of the computational graph represent quantum logic
gates that describe interactions between qubits. 
The final vertices of the graph correspond to output states.
Infinite computations correspond to graphs that need not have 
final states.

The quantum computation gives rise to an amplitude
${\cal A} = \langle 00 \ldots 0| U |00\ldots 0\rangle$,
where  $|00\ldots 0\rangle$ is the initial state of the qubits, 
$U=U_n \ldots U_1$ is the unitary operator given by the product
of the unitary operations $U_\ell$ corresponding to the individual quantum
logic gates that act on the qubits one or two at a time, and 
$\langle 00 \ldots 0|$ is the final state (figure 1).
Infinite computations do not possess an 
overall amplitude, but still assign conditional amplitudes 
to transitions between states within the computation.

\bigskip\noindent{\it Example: quantum simulation}

As noted above,
quantum computation can reproduce the behavior of any discrete
quantum mechanical system [2-6].  Let's review how a quantum computer
can reproduce the dynamics of a quantum system, such as a lattice
gauge theory, whose dynamics consists of Hamiltonian interactions 
between discrete degrees of freedom.   Consider
a Hamiltonian of the form $H=\sum_\ell H_\ell$, where each Hamiltonian
$H_\ell$ acts on only a few degrees of freedom, e.g., fields
at neighboring lattice sites.  Using the
Trotter formula, the time evolution of this Hamiltonian
can be written 
\begin{eqnarray}
e^{-iHt} = e^{-iHt/n} \ldots e^{-iHt/n}
= \Pi_\ell e^{-iH_\ell t/n} \ldots
\Pi_\ell e^{-iH_\ell t/n} + O(\|Ht\|^2/n.
\end{eqnarray}
so that the overall time evolution can be approximated
arbitrarily accurately using a sequence of local transformations
simply by slicing time finely enough [3]. 
Now each $e^{-iH_\ell t/n}$ can be enacted using only a small
number of quantum logic gates:
$e^{-iH_\ell t/n} \approx  U_m \ldots U_1$.
Here the number of logic gates $m$ required to approximate
$e^{-iH_\ell t/n}$ to accuracy $\epsilon$ goes as $\epsilon^{-d^2}$,
where $d$ is the dimension of the local Hilbert space acted on by $H_\ell$.
(For example, for pairwise interactions between two-level 
systems, $d=4$.)
The logic gates $U_k$ used to enact the infinitesimal time
evolutions $e^{-iH_\ell t/n}$ can themselves involve only
infinitesimal rotations in Hilbert space.  

Accordingly, any local Hamiltonian dynamics, including, e.g., lattice
gauge theory, can be reproduced to any desired degree of accuracy 
using a sequence of infinitesimal quantum logic operations [3]. 
Indeed, a lattice gauge theory can be thought of as a special case
of a quantum computation, in which the quantum degrees of freedom
are the quantum fields at different lattice points,
and the quantum logic gates are infinitesimal Hamiltonian
interactions coupling fields at the same point or at neighboring
points.  Fermionic systems can be reproduced either by quantum computations
that use fermionic degrees of freedom [5], or by local interactions
that reproduce fermionic statistics [4,6,31].   As a result,
the techniques given here to derive quantum gravity from the
sequence of quantum logic operations in a quantum computation 
will serve equally well to derive a theory of quantum gravity from 
the sequence of infinitesimal Hamiltonian interactions in
a lattice gauge theory such as the standard model.  In the
former case, the concept of distance is derived from the interactions
between qubits; in the latter, it is derived from the interactions
between quantum fields.

\subsection{Computational histories}

In the computational universe, the structure of spacetime
is derived from the behavior of quantum bits as they
move through the computation.
At each vertex of the computational graph, depending on the
state of the incoming quantum bits, those qubits can either be transformed 
(scattering), or not (no scattering).  When qubits scatter,
that constitutes an event.  If the bits are not transformed 
(no scattering), then there is no way to tell that they have
interacted: no scattering, no event.  
Each computation is a superposition
of different computational histories, one for each pattern of 
scattering events (figures 2 and 3).

A scattering/no-scattering superposition in
the computation corresponds to a fluctuation in the
path that information takes through the computation.  
Because such an event/no-event superposition is a 
superposition of different causal structures, it will be seen
below to correspond to a fluctuation in spacetime geometry. 
To make this `scattering -- no scattering' picture explicit, 
consider quantum logic gates of the form
$U= e^{-i\theta P}$, where $P^2=P$ is a projection operator.
In this case, $P(1) \equiv P$ projects onto the eigenspace
of $U$ with energy $1$, and $P(0) \equiv 1-P$ projects onto
the eigenspace with energy $0$.  $U$ can be written as
$U= P(0) + e^{-i\theta} P(1)$: states in the $0$ eigenspace of $P$
do not interact (no scattering), while states in the $1$ eigenspace interact
and acquire a phase $e^{-i\theta}$ (scattering).
For example, if $P$ is the projector onto the two-qubit triplet subspace,
then $U= e^{-i\theta P}$ continuously `swaps' the input qubits; such
transformations are universal on a subspace of Hilbert space [32-33].  

The generalization to quantum logic gates with more than one non-zero
eigenvalue is straightforward (see Methods A1.2): in this case,
each non-zero eigenvalue corresponds to a scattering event, but
different eigenvalues gives rise to a different phase
associated with the vertex.  If there are
no non-zero eigenvalues, then all computational histories have
the same causal structure, but differ in the action associated
with each vertex.

The vertices of a computational history correspond to scattering events.  
A computation with $n$ logic gates 
gives rise to $2^n$ computational histories $C_b$, where
$b = b_1 \ldots b_n$ is an $n$-bit string containing a 1 for each scattering
event and a 0 for each non-scattering non-event.  
The overall unitary transformation for the computation,
$U = U_n\ldots U_1$, can be decomposed in to a sum over 
the $2^n$ computational histories $C_b$:
\begin{eqnarray}
 U_n\ldots U_1 =
(P_n(0) + e^{-i\theta_n}P_n(1)) \ldots (P_1(0) + e^{-i\theta_1}P_1(1)) 
\nonumber \\
= \sum_{b_n\ldots b_1 = 00\ldots0}^{11\ldots 1}
e^{-i\sum_\ell b_\ell \theta_\ell} P_n(b_n) \ldots P_1(b_1). 
\quad\quad\quad\quad(1.1)
\end{eqnarray}
Each computational history $C_b$ has a definite phase $e^{-i\theta_\ell}$
associated with each scattering event $b_\ell = 1$ (figures 2 and 3).
Here, $\theta_\ell$ is the phase acquired by the nonzero-energy eigenstate 
of the $\ell$'th quantum logic gate: $\theta_\ell$ is an angle rotated
in Hilbert space.  Define the {\it action} of a computational
history $C$ to be $I =  \hbar \sum_{\ell \in v(C)} \theta_\ell,$ where 
$v(C)$ are the vertices of $C$.  If we think of each quantum logic gate
as supplying a local interaction at a scattering event, then
$\hbar\theta_\ell$ is equal to the energy of interaction times the
time of interaction.  Because in quantum mechanics only {\it relative}
phases are observable, we take the lowest energy eigenstate
to have eigenvalue 0, and the phases $\theta_\ell$ to be positive.  

A quantum computation is a superposition of computational histories.
As will now be shown, each computational history corresponds to a 
discrete classical spacetime with a metric that obeys the discrete
form of Einstein's equations.  The superposition of computational
histories then gives rise to a superposition of classical spacetimes,
like the superposition of paths in a path integral.  

\section{General relativity and Regge calculus}

Einstein derived the theory of general relativity from the
principle of general covariance [34]:  the laws of gravitation 
should take the same form no matter how one chooses to assign
coordinates to events.
To relate quantum computation to general relativity, embed the
the computational graph in a spacetime manifold by mapping
$C$ into $R^4$ via an embedding mapping ${\cal E}$. 
Vertices of the embedded graph correspond
to events, and wires correspond to the paths information takes in
the spacetime.  The embedding mapping should respect the
causal structure of $C$: because it is a directed, acyclic
graph, $C$ contains discrete analogs of Cauchy surfaces, sets of  
points which non-extensible causal paths intersect exactly once.    
The embedding should map each discrete Cauchy set of $C$ 
into a Cauchy surface of the spacetime; and a foliation
of $C$ in terms of discrete Cauchy sets should be mapped
into a foliation of the spacetime in terms of Cauchy surfaces.

The geometry of the spacetime is derived
from the embedded computation.  The information that moves
through the computation effectively `measures' distances in 
spacetime in the same way that the signals passed between
members of a set of GPS satellites measure spacetime. 
The observational content of a quantum computation consists
of its causal structure together with the local action of the
computation.  Each computational history gives rise to a particular
causal structure, embodied in the history's directed graph,
and a particular phase or action associated with each vertex
of that directed graph.  All other quantities, such as edge lengths,
curvature, and the stress-energy tensor, are to be deduced from
this causal structure and action. 
   
Since the way that information is processed in a quantum computation
is independent of the way in which that computation is embedded 
in spacetime, any dynamical laws
that can be derived from the computation are automatically generally 
covariant and background independent: 
the observational content of the theory (causal structure
and action) is invariant under general coordinate
transformations.  Since general covariance (together with restrictions
on the degree of the action) implies Einstein's equations,
the geometry induced by the computational universe obeys
Einstein's equations (in their discrete, Regge calculus form [35-37]).
We now verify this fact explicitly.

\subsection{Regge Calculus} 

Because the computational graph is at bottom a lattice picture
of spacetime, the computational universe is based on the Regge calculus 
version of general relativity [35-37].  In Regge calculus, spacetime
geometry is defined by a simplicial lattice whose edge lengths
determine the metric and curvature.  This simplicial lattice
is a discrete, `geodesic dome' analog of the manifold.  
Each computational history $C_b$ gives rise to a Regge calculus
by the following steps.  The computation as a whole corresponds
to a superposition of Regge calculi, one for each computational history.

The lattice determined by a given computational history $C$ 
is not itself simplicial.  Extend it to a simplicial Delaunay
lattice $D_C$ by adding edges and vertices [38].  
All edges of the computational history are edges of the simplicial
lattice (they may be divided into several segments by additional
vertices).  The four edges at each vertex of the computational graph,
two incoming and two outgoing, define a four-simplex associated
with each vertex: the pairwise nature of interactions between
quantum degrees of freedom gives rise naturally to a four-dimensional
discrete geometry [figure 4].    

There are a variety
of methods for extending a lattice to form a simplicial lattice
[38]; two are described in the Methods.  
{\it No matter how one constructs
such an extension, the resulting discrete spacetime still possesses
the same observable content.}  The causal structure and action of 
the discretized spacetime depend only on the computational history
$C$, and not on the choice of simplicial lattice used.  
That is, the microscopic details of $D_C$ do not
affect the observable structure of spacetime.  Just as two surveyors
can triangulate the same landscape in different ways, and yet
obtain the same geometry for the region, so two different
simplicial triangulations of the computational history possess
the same causal structure and action.  Consequently,
as will be seen below, different triangulations also give
rise to the same curvature associated with a vertex. 

The simplicial lattice $D_C$ gives the edges of 
a four-dimensional geodesic dome geometry, and its dual 
lattice $V_C$ defines the volumes $\Delta V_\ell$ 
associated with the vertices.   The simplicial lattice corresponds
to a four-dimensional Lorentzian spacetime [35-37].
In the usual construction of Regge calculus, the variables that
define the geometry are the lengths of the edges
of the simplicial lattice.  In the computational theory considered
here, these lengths are defined by the
dynamics of the computation.  Along wires, no phase is accumulated
and the action is zero.  Accordingly, identify these wires with
null lines $E_a$, for $a=1$ to $4$.  

Embedding a vertex and its neighbors in $R^4$ and using
these four null lines $E_a$ as a null basis implies that the on-diagonal
part of the metric is zero: $g_{aa} = g(E_a, E_a) = 0$.    
(If the $E_a$ are not linearly independent, almost any small
change in the embedding will make them so.)
The off-diagonal part of the metric is to be determined by a 
requirement of self-consistency.  Once one picks the off-diagonal parts of 
the metric at each vertex, the
length of each edge is fixed by averaging the lengths given by
the metrics on the two vertices associated with the edge.
That is, the length of the edge $l$ connecting
vertices $j$ and $k$ is equal to 
$g_{ab}(j) l^al^b/2 + g_{ab}(k) l^al^b/2$. 
This definition of edge length is covariant. 
To enforce the requirement of self-consistency, assume that the full metric 
has been chosen at each point, so that edge lengths and the full geodesic 
dome geometry are fixed.  As in the reconstruction of the discrete
metric from GPS data, the causal structure and local action will
then determine a self-consistent choice of the off-diagonal terms 
of the metric.

Once the off-diagonal terms have been chosen, Miller's elegant
construction of the connection and curvature tensor for Regge calculus
on Delaunay and Voronoi lattices can be applied [36].
In four dimensional Regge calculus, curvature lies on triangular hinges.
Once the edge lengths have been defined,
each hinge has a well-defined curvature, volume, and action associated with
it.  The gravitational
action corresponding to a hinge is just $(1/8\pi G) A \epsilon$,
where $A$ is the area of the hinge and $\epsilon$ is the deficit angle
of the hinge.  The action corresponding to a hinge is invariant under coordinate
transformations.

\subsection{The Einstein-Regge equations}

The Einstein-Regge equations are obtained by demanding that the
combined action of gravity and matter are stationary under variations
of the metric [39].  The usual way of deriving the Einstein-Regge
equations is to demand that that the action be stationary under
variations in the individual edge lengths: $\delta I/\delta l(jk) =0$.
The edge lengths determine the metric, and {\it vice versa},
so stationarity under variations in the edge lengths is
equivalent to stationarity under variation in the metric, 
$\delta I / \delta g_{ab}(j) = 0$. 

The action of the gravitational degrees of freedom
in Regge calculus is a scalar quantity given by
the sum of the actions over the individual
hinges: $I_G = (1/8\pi G)\sum_h A_h \epsilon_h$, where $G$ is
the gravitational constant.  
For finite computations this action should be supplemented
with a term, $(1/8\pi G) \sum_k  B_k \beta_k$,
corresponding to the extrinsic curvature of the three dimensional
boundary of the computation [40-41].  Here,
$B_k$ is the area of the $k$'th triangle joining two tetrahedra
at the boundary of the lattice, and $\beta_k$ is the angle between
the normals to those two tetrahedra. 
$A_h$ and $\epsilon_h$ are functions of the metric
at that point and at neighboring points.  Define the action of
the computational matter to be the scalar quantity
$I_M = - \hbar \sum_{\ell \in v(C)} \theta_\ell $,
proportional to the overall action $I$ of the computational
history. The Einstein-Regge equations at a point $\ell$ are then given by
\begin{eqnarray}
\delta I_G/ \delta g_{ab}(\ell) + 
\delta I_M/ \delta g_{ab}(\ell) = 0 .
\end{eqnarray}
In the Lagrangian approach to general relativity 
$\delta I_G/\delta g_{ab}(\ell)$ is equal to
$(-1/16\pi G) \Delta V_{\ell}$
times the discretized Einstein tensor, $G^{ab} = R^{ab} - (1/2) g^{ab} R$;
similarly, $\delta I_M/\delta g_{ab}(\ell)$ is equal to
$(1/2) T^{ab} \Delta V_\ell$ where $T^{ab}$ is the energy-momentum tensor
([39] section 3.3).
Here, $\Delta V_\ell$ is the Voronoi volume at point $\ell$.

To obtain an explicit form for the Einstein-Regge equations,
write the action for the computational matter $I_M = 
-\hbar \sum_{\ell\in v(C)}\theta_\ell$ as 
$\sum_\ell {\cal L}_\ell {\Delta V_\ell} $
where ${\cal L_\ell}$ is the local Lagrangian at point $\ell$. 
Following Einstein and taking the Lagrangian to be a function
only of the computational `matter' together with the metric to
first order, we have
\begin{eqnarray}
{\cal L}_\ell = - g_{ab} \hat T^{ab}/2 - {\cal U}_\ell,
\end{eqnarray}
where $\hat T^{ab} = {\rm diag}(\gamma_1, \gamma_2,\gamma_3,\gamma_4)$ 
is diagonal in the null basis, so that $g_{ab} \hat T^{ab} = 0$.
Here, ${\cal U}_\ell$ is the energy density of the interaction between
the qubits at the $\ell$'th quantum logic gate; it gives
the potential energy term in the energy momentum tensor.
As will be seen below, $\hat T^{ab}$ represents the 
kinetic energy part of the energy momentum tensor, and
$\gamma_a$ is the energy density of the `particle' moving
along the vector $E_a$.  $\hat T^{ab}$ is a traceless tensor that
does not contribute to the action, but does
contribute to the energy-momentum tensor.  For the moment,
the $\gamma_{a}$ are free parameters: since the choice of the
$\gamma_{a}$ does not affect the action, they can be chosen 
in a suitable way, e.g., to make the Einstein-Regge equations hold.

The computational `matter'
looks like a collection of massless particles traveling along
null geodesics and interacting with each other via
two-particle interactions at the quantum logic gates.
$\hat T^{ab}$ is the kinetic energy term for
these massless particles, and ${\cal U}$ is the
potential energy term.  This relation of this `quantum computronium'
to conventional quantum fields will be discussed below.
Note that ordinary matter seems to have ${T^a}_a \leq 0$, 
corresponding to non-negative energies and phases 
$\theta_\ell$ for the quantum logic gates.

The Einstein-Regge equations (2.1) now 
take the explicit form (Methods A2.2):
\begin{eqnarray}
-\sum_{h\in N(\ell)} 
\epsilon_h {\delta A_h \over \delta g_{ab}(\ell)} 
= 4\pi G T^{ab} \Delta V_\ell 
\end{eqnarray}
where the energy-momentum tensor
$ T^{ab} =  \hat T^{ab}(\ell) - {\cal U}_{\ell} g^{ab}(\ell) $. 
Here, the sum over hinges includes only hinges $h\in N(\ell)$ 
that adjoin the vertex $\ell$, as the edges of other hinges do not change
under the variation of the metric at $\ell$.

Equations (2.3) are the form for the Einstein-Regge 
equations in the computational universe (they can also be
written in terms of edge lengths and angles as in [35], Methods A2.2).   
Just as in the ordinary Einstein equations, the left-hand side
of (2.3) contains only geometric terms, while the right-hand side contains
terms that depend on the action of the computational matter and
on the metric to first order.  
The terms on the left hand side of equation (2.3) are functions
of $g_{ab}$ at $\ell$ and at adjacent points.  The right hand side
of (2.3) depends only on the computational `matter.'
In other words, equation (2.3) is a nonlinear partial difference equation
that relates the curvature of space to the presence of matter, with the
local phase of the computational matter acting as a source for
curvature.  The computational matter tells spacetime how to curve.

Now we can verify that the gravitational action associated with a
vertex is independent of the embedding and simplicial extension.  
Taking the trace of equation (2.3) yields
$R_\ell \Delta V_\ell = 32\pi G\hbar\theta_\ell$, where $R_\ell 
\Delta V_\ell/16\pi = I_{G\ell} = 2\hbar\theta_\ell$ 
is the action of the gravitational field
associated with the point $\ell$.  In other words, like the action
of the computational `matter,' the gravitational
action is just proportional to the angle rotated at the vertex;
it is a scalar quantity that is invariant under embedding and 
simplicial extension.  Up until now, we have been associating
the computational action at each vertex with the action of
the `matter' fields alone.  However, since $I_{G\ell} =
2\hbar\theta_\ell$, we can associate the computational action
with {\it both} matter and gravity.  Indeed, the total action
of matter and gravity associated with the $\ell$'th vertex
is just the computational action 
$ I = I_{M\ell} + I{G\ell} = -\hbar\theta_\ell + 
2 \hbar\theta_\ell = \hbar\theta_\ell$. 
The dynamics of the quantum computation give rise both to matter and
to gravity, coupled together. 

As in [35], when the average curvature
of the simplicial lattice is small, corresponding either to 
length scales significantly larger than the Planck scale,
or to small angles rotated at each vertex, 
the ordinary Einstein equations can be obtained from equations
(2.3) by a process of coarse-graining (Methods section A2.2-3).  
Note that in the example given above of a quantum computer reproducing
the dynamics of a system with a local Hamiltonian such as lattice
gauge theory, the angles rotated at each vertex are indeed small, 
so that in this case coarse-graining may be expected to give an 
accurate reproduction of the underlying discrete dynamics. 

\subsection{Satisfying the Einstein-Regge equations}

The Einstein-Regge equations (2.3) govern the self-consistent
assignment of lengths in the simplicial geometry.
In order to complete the construction of discrete geometry from 
an underlying computational history, the parameters that so far have 
remained free must be fixed.  These parameters are the off-diagonal part of
the metric, and the on-diagonal part of the energy-momentum tensor.  
Self-consistency dictates that these parameters be chosen in
such a way that equations (2.3) are satisfied.

Equations (2.3) express the relationship between matter and geometry
for one computational history in the discrete, computational universe.  
They are a set of ten coupled, nonlinear
difference equations which can be solved (e.g., numerically)
given appropriate boundary conditions.  To fix the free parameters, 
first choose the six off-diagonal terms in the metric to satisfy 
the off-diagonal part of equation (2.3).  Next, choose
the four on-diagonal terms in the energy-momentum tensor,
$\gamma_{a}$, so that remainder of the Einstein-Regge equations
(2.3) are obeyed.   The four on-diagonal
terms $\gamma_a$ correspond to the `kinetic energies' of the four 
qubits as they move through the computation.  
Essentially, choosing $\gamma_{a}$ in this fashion
reduces to insisting that energy and momentum travel along the null
geodesics specified by the causal structure, and demanding that
${T^{ab}}_{;b}=0$.   The traceless part of the
gravitational terms in the Einstein-Regge equations can be thought
of as source terms that determine the kinetic energy of the computational
matter: spacetime tells qubits where to go.  

At reference vertices and at vertices corresponding to single-qubit
logic gates, fewer null lines are determined by the computational
history.  As shown in Methods, section A2.3,
the Einstein-Regge equations can still be used to satisfy
the self-consistency conditions at such vertices.

This completes the construction: each computational history gives rise 
to a discrete geometry that obeys the Einstein-Regge equations.

\subsection{2.4 Minimum length scale is the Planck scale}

Because only relative phases are observable, we have adopted
the convention that local phases and curvature are positive:
$\epsilon_h > 0, \theta_\ell > 0$.  Consequently, a minimum
length scale arises because the maximum positive deficit
angle is $\epsilon_h = 2\pi$.  In equation (2.3), the
curvature terms are on the order of $\epsilon_h A_h$, while
the minimum phase $\theta_\ell$ required to flip a qubit is
$\pi$.  As a result, if area of a hinge $A_h$ adjacent to an
event in which a bit flips is less than the Planck area $ \approx\hbar G$, 
it is not possible to satisfy (2.3):
space would have to curve more than it can curve.

If we relax the assumption of positive phases and curvature, 
($\epsilon_h < 0, \theta_\ell < 0$),
this argument does not seem to apply, as there is no limit on
how negative the deficit angle can be.  As noted above, however,
ordinary matter seems to have
${T^a}_a \leq 0$, corresponding to non-negative local phases, energies,
and curvatures.

\section{Observational consequences of
the computational universe.}

The previous section showed how each computational history
gives rise to a classical discrete spacetime that obeys the
Einstein-Regge equations.  As shown in section (1), a quantum computation
is a superposition of computational histories, and so gives
rise to a superposition of spacetimes.  Now investigate the
implications of the computational universe picture for several 
well-known problems in quantum gravity.

\subsection{The back reaction}

The first problem for which the computational universe provides an
explicit solution is the back reaction of the metric to
quantum-mechanical matter [7].  Here, since the metric is derived
from the underlying dynamics of quantum information, its fluctuations
directly track the quantum fluctuations in the computational
matter.  In particular, fluctuations in the curvature scalar track
the fluctuations in the local phase or action accumulated by the
computational matter; and quantum fluctuations in the routing of signals
through the computation (via scattering or lack or scattering) give
rise to fluctuations in the remainder of the metric.  For example,
such fluctuations can give rise to gravity waves via the usual
mechanism in which the Ricci tensor and curvature scalar act as
source terms for gravity waves in classical relativity via the
Bianchi identities ([39] section 4.1).

The computational universe model is intrinsically a theory of 
quantum matter {\it coupled} to gravity, and not a theory of
either quantum matter or quantum gravity on its own.  But it
still supports such `purely gravitational' effects as gravity waves.
A computation that contains a binary pulsar will also contain the
gravity waves emitted by that pulsar.  Quantum fluctuations in
the positions and momenta of the two stars that make up the pulsar
will give rise to fluctuations in the gravity waves emitted.
Those gravity waves will
propagate through the computational vacuum and will induce 
disturbances in the computational cosmic background radiation; finally, 
if the computation contains LIGO, it will detect those gravity waves.

The intertwined nature of quantum matter and quantum metric in this
theory implies that when the matter decoheres (see section (3.5) below), so
does the metric.  But the metric does not independently act as a source
of decoherence for the underlying quantum mechanical matter (in 
contrast with [11, 42]).  Consequently, the experiment proposed
in [42] should reveal no intrinsic decoherence
arising from the self-energy of the gravitational interaction.

\subsection{Singularities and black hole evaporation}

Initial singularities in the computational universe correspond
to places in the computation where bits are added to the computation,
as in section (1.1) above.   As soon as a newly minted bit interacts
with another bit, a new volume of spacetime is created; its size
is determined by solving equation (2.3).
Similarly, final singularities occur when bits 
leave the computation, corresponding to a projection onto a final bit,
also as in (1.1).  When the bits go away, so do the volumes of spacetime
associated with them.  Initial and final singularities are essentially
the time reverse of each other; they are quite `gentle' compared with
our normal view of such singularities.  Although the energy scales
associated with such singularities can be high, the process
of bits entering and leaving the computation is orderly and
quantum-mechanically straightforward.

Singularities obey a version of the cosmic censorship hypothesis.
Final singularities occur at points where bits go away: there
are no future directed `wires' leading away from them.  Similarly,
initial singularities occur at points where bits come into existence:
there are no past directed wires leading into them.

The behavior of the computation at final singularities gives
rise to a mechanism for information to escape from black holes.
In a finite computation with 
projection onto final states $\langle 00\ldots 0|$, black hole evaporation 
can take place by a variant [43] of the Horowitz-Maldacena mechanism [44].
As shown in [43-47], some, and typically almost all, of the quantum 
information in a black hole escapes during the process of evaporation.
If microscopic black holes can be created experimentally, e.g., in
high-energy collisions, the theory predicts that their evaporation 
should be approximately unitary.

\subsection{Holography and the geometric quantum limit}

In the case that local energies and phases are taken to be positive,
the geometric quantum limit [48]
bounds the number of elementary events (ops)
that can be fit into a four volume of spacetime to $\#
\leq (1/2\pi) (x/\ell_P)(t/t_P)$, where $x$ is the spatial extent
of the volume and $t$ is the temporal extent.  The geometric
quantum limit is consistent with and implies holography [48-51].
A detailed description of the geometric quantum limit
and its relation to holography can be found in reference [48].

Note that the geometric quantum limit implies that the separations
between bits and quantum logic operations within a volume of space time
are typically much greater than the Planck length.  In particular,
for our universe as a whole, the total number of ops and bits is
less than or equal to $t^2/t_P^2 \approx 10^{122}$.  This yields
an average spacing between ops in the universe up to now 
of $\sqrt{tt_P} \approx 10^{-13}$ seconds.  (As will be seen
below, this time scale can be related to the mass scale for
fundamental spinors.)
Of course, here on Earth, where matter is packed more densely than in
the universe as a whole, bits and ops are crammed closer together.

\subsection{Quantum cosmology}

The Einstein-Regge equations (2.3) can be solved under certain
simplifying assumptions (e.g., large computations with random wiring
diagrams, or uniform, cellular-automaton type architectures) to calculate
the spectrum of curvature fluctuations in the early universe [52].
The approximate spatial homogeneity of random and CA-like architectures
gives an approximately homogeneous, isotropic universe in 
which the Einstein-Regge
equations take on a Friedmann-Robertson-Walker form.
Note that cellular automata may not be isotropic: particular
directions may be picked out by the axes of the automaton.
However, the example of lattice gases shows that such architectures
can still give rise to an isotropic dynamics at a coarse-grained
level.  In particular, if the underlying dynamics is that of
lattice QCD, we expect the coarse-grained dynamics to be
isotropic. 

For homogeneous, isotropic spacetimes, 
the coarse-grained metric (averaging over microscopic
fluctuations) takes on the form [39] 
\begin{eqnarray}
{\rm d}s^2 = - {\rm d}t^2 + a^2(t)\big({\rm d}\chi^2 + 
f^2(\chi)({\rm d} \theta^2 + \sin^2\theta {\rm d} \phi^2) \big). 
\end{eqnarray}
Here $f(\chi) = \sin \chi$ for three spaces of constant positive
curvature, $f(\chi) = \chi$ for three spaces of zero curvature,
and $f(\chi) = \sinh \chi$ for three spaces of constant negative 
curvature.  The homogeneity of the spacetime requires the 
energy momentum tensor of the matter to take on the form
of a perfect fluid, with energy density $\rho$ and pressure
$p$.  The Freedman-Robertson-Walker (FRW) equations then take on the form
\begin{eqnarray}
\dot\rho = -3(\rho + p) \dot a/a  \\  
4\pi G(\rho+3p)/3 = - \ddot a/a  \\
8\pi G \rho/3 = \dot a^2/a^2 - k/a^2, 
\end{eqnarray}
where $k=-1, 0, +1$ for three spaces of positive curvature, zero curvature,
and negative curvature, respectively.  The third of
these equations is just the integral of the first, so there
are really only two independent FRW equations.

Recall that our prescription is first to assign the free parameters
of the metric so that the off-diagonal, potential energy ($U$)
parts of the Einstein-Regge equations hold, and then to assign
the kinetic energy terms so that the on-diagonal parts of the
equations hold.  Under the assumptions of homogeneity and
isotropy, the potential energy part of the energy-momentum
tensor takes on the form $diag( U,-U,-U, -U)$, while the
(traceless) kinetic energy part takes on the form
$diag (K, K/3,K/3,K/3)$, so that $\rho = K + U$, 
and $p= K/3 -U$.  Defining the Hubble parameter $H= \dot a/a$,
we can rewrite the FRW equations (3.2) as
\begin{eqnarray}
16\pi G U /3 = \dot H + 2H^2 (3.3a) \\
8\pi G(K+U) /3 =& H^2 - k/a^2. (3.3b)
\end{eqnarray}
Recall that the potential associated with a four volume of spacetime
 $\Delta V$ is $U = \hbar \theta /\Delta V$, where $\theta$ is the net
phase acquired within $\Delta V$.  Fixing $a$ fixes the scale of
the volume $\Delta V$.  Accordingly, for a given underlying computational
geometry, equation (3.3a) can be regarded as a differential equation in $a$.  
Once this equation is solved, the kinetic energy $K$ can be
assigned according to equation (3.3b).  

Let's look at some different possible behaviors
of the FRW equations in the computational universe.  The FRW
equations can conveniently be rewritten as
\begin{eqnarray}
-16\pi G K /3 = \dot H (3.4a)\\
8\pi G(K+U) /3 =& H^2 - k/a^2. (3.4b)
\end{eqnarray}
First, consider the case of a flat universe, $k=0$.
The case of positive intrinsic curvature, $k=-1$, will
be investigated below.
(Note that the positivity of energy requires $H>1$ for
a universe with negative curvature $k=1$.)
We can distinguish between several different regimes.  

\smallskip\noindent
{$\bullet$} First,
take the case $K=0$.  Then $\dot H =0$, $H$ equals a
constant, as does $U$, and the universe undergoes
inflation at a constant rate: $a(t) = a(0) e^{\alpha t}$.

\smallskip\noindent
{$\bullet$} If $U>>K$, then the universe also expands exponentially;
but as long as $K>0$, equation (3.4a) shows
that the rate of expansion to decreases over time.
Since $\ddot a/a = 8\pi G (U-K)/3$, 
when $K>U$, then $\ddot a < 0$, and the universe ceases
to inflate.

\smallskip\noindent
{$\bullet$} The situation $K>>U$ 
corresponds to a radiation dominated universe, 
and equations (3.4) have the solution $a(t) \propto t^{1/2}$.  

\smallskip\noindent
{$\bullet$} The situation $K\approx 3U$
corresponds to a matter dominated universe ($p= K/3 -U \approx 0$),
and equations (3.4) have the solution $a(t) \propto t^{2/3}$.

All of these scenarios are possible depending on the
underlying computation; and each of these scenarios can
hold at different stages in the same computation.
For example, consider a homogeneous, isotropic computation
commencing at $t=0$.  In order to solve equation (3.3a),
initial conditions on $a$ and $K$ must be set.  A natural
choice is $a=1$ (in Planck units) and $K=0$.  These initial
conditions correspond to the first scenario above,
inflation at the Planck rate.  
Just as in conventional inflation, microscopic fluctuations in local phases
give rise to Gaussian curvature fluctuations at the coarse-grained 
scale, which in turn should be inflated as in the conventional
picture of the growth of quantum-seeded curvature fluctuations.

Now equation (3.4a) shows that this inflation 
is unstable: in any region where
the kinetic energy is not strictly zero, the rate
of inflation decreases.  The greater the kinetic energy in
some region, the more rapid the slowdown.  In regions where  
$K$ grows to be greater $U$, $\ddot a$ becomes less than $0$ and
inflation stops.  Where does this kinetic energy come
from?   At microscopic scales,
homogeneity is broken by quantum fluctuations in the phases
$\theta$ acquired in individual gates.  Accordingly,
even if the kinetic energy terms in the energy-momentum
tensor are initially zero everywhere,
the full, microscopic solution of equations (2.3) will
give rise to kinetic energy locally.   Inflation will
then slow, and energetic matter will be created.  In
other words, just as in the standard picture of inflation,
Planck-scale inflation is unstable to the creation
of ordinary matter, which nucleates a phase transition
and gives rise locally to a radiation-dominated universe.

Uncovering the details of this phase transition will require
the detailed solution of equation (2.3), and is beyond the
scope of the current paper.  Within the confines of
the assumption of homogeneity
and isotropy, the FRW equation nonetheless can provide useful 
qualitative information about the behavior of computational
cosmologies.   For example, consider a radiation dominated
universe such as the one created after inflation in the
previous paragraph.  The kinetic energy density in such
a universe drops over time as $K \propto 1/a^4 \propto t^{-2}$,
while the potential energy density $U$ is proportional to
the density of logic gates and goes as $U \propto 1/a^3
\propto t^{-3/2}.$  Eventually $K$ lowers to the level
$K=3U$ and the universe becomes matter dominated. 
Not surprisingly, since a homogeneous computation such as
a quantum cellular automaton can give
rise to conventional matter and energy, and since
a homogeneous, isotropic computational
universe obeys the FRW equations, the usual picture of a
radiation-dominated universe giving way to a matter-dominated
one holds in the computational universe.

By the time radiation domination has given way to matter
domination, the universe is no longer homogeneous: significant
clumping has taken place.  Therefore, in addition to matter-dominated
regions, we expect some regions to have $U<K<3U$.  These regions
exhibit negative pressure $p=K/3 -U$, but do not have
$\ddot a >0$: the negative pressure is not great enough
for them to inflate.  Depending on the scale
of the inhomogeneities, other regions may have $K<U$.  As
(3.2b) shows, these
regions will begin to undergo inflation again, though at
a rate much smaller than the Planck scale. 

We see that at this late date, the computational universe can contain
regions dominated by three different types of energy: (1) ordinary matter and
radiation, $K\geq 3U$, (2) matter (cold dark matter?) with negative 
pressure that does not inflate, $U\leq K<3U$, and (3) `dark energy'
that undergoes inflation, $K<U$.  The relative preponderance
of these regions depends on how inhomogeneities develop in the
full Einstein-Regge equations (2.3), and lies beyond the scope 
of this paper.  From continuity, we expect that matter of
type 2 should lie in regions such as the halos of galaxies
intermediate between ordinary matter (type 1) and dark energy (type 3)
(this is the reason for the speculative identification of matter
of type 2 with cold dark matter).  In accordance with equation (3.3b), 
in a flat universe, the total density of all three types of matter is equal
to the critical density $3H^2/8\pi G$.  That is, in a universe dominated
by dark energy after the radiation and matter-dominated epochs, 
the rate of inflation is determined by the Hubble parameter at that epoch.
In this picture, we see that primordial inflation and late-epoch
reinflation are caused by the same mechanism, a locally potential
dominated universe.  Because length scales are dynamically determined,
however, the rate of late-epoch inflation can be many orders
of magnitude (e.g. $10^{60}$ times) smaller than the rate
of primordial inflation.

\subsection{Universes with positive intrinsic curvature}

Let us look at a specific example of a FRW cosmology
in the computational universe.
In our model, curvature is non-negative.  If the underlying
computation has approximately uniform distribution of
quantum gates, then we are led to consider models with
constant positive curvature, i.e., de Sitter space [39].
De Sitter space has topology $R^1 \times S^3$ and
can be visualized as a hyperboloid
$-v^2 + w^2 + x^2 + y^2 + z^2 = \alpha^2$
embedded in flat $R^5$, with metric
$  {\rm d}s^2=-{\rm d} v^2 + {\rm d}w^2 + {\rm d}x^2 + {\rm d}y^2 + 
{\rm d}z^2  $.  We can arrange gates uniformly
on this hyperboloid with wires following null lines
of the metric.  
The constant curvature of de Sitter space is $R = 12/\alpha^2$.
The energy momentum tensor is purely potential: $K=0$,
$U=R/32\pi G$.   Consider gates that correspond to swaps or to
bit flips, so that $U\Delta^4 V = \pi\hbar/2$.  Since
$U= 3/8\pi G \alpha^2$, the spacing between gates
is $(\Delta^4 V)^{1/4} = (4\pi/3)^{1/4} \sqrt{t_P \alpha}$.

The number of bits on successive
three-spheres in the $v$ direction grows as
$(\alpha^2 + v^2)^{3/2}$, so this computation involves
continual bit creation for positive $v$, and continual
bit distruction for negative $v$.  The positive $v$ and
negative $v$ parts of the computation can be time reversals
of eachother.  Note that the
discrete nature of the underlying computation
means that it is not necessary for the positive $v$
part of the computation on the hyperboloid to be causally connected with
the negative $v$ part: there need not be any `wires' passing
through the waist at $v=0$.  In this case, the $v\geq 0$
part of the computation can be taken as a model for our
universe starting from a big bang.

Setting $\alpha \sinh(t/\alpha) = v$ and using angular 
coordinates $\chi, \theta, \phi$ for the three sphere
gives the metric in FRW form for closed spacelike three
spaces ($k= -1$) [39]:
\begin{eqnarray}
{\rm d} s^2 = -{\rm d}t^2 + \alpha^2 \cosh^2(t/\alpha)
\big( {\rm d}\chi^2 + \sin^2 \chi ( {\rm d}\theta^2 
+ \sin^2\theta {\rm d} \phi^2) \big).  
\end{eqnarray}
De Sitter space expands exponentially at a rate
determined by $\alpha$.  

The qualitative discussion of flat FRW spaces given also holds
for this de Sitter space computation.  In particular,
suppose that $\alpha$ is taken to be equal to 1,
yielding Planck scale inflation. If the distribution
of phases rotated in the gates were exactly uniform,
then this inflation would continue for ever,
and the de Sitter picture would be essentially exact.
However, just as before, Planck scale inflation is
unstable to the creation of kinetic energy,
which slows the rate of expansion.  In other 
words, $\alpha$ cannot be a constant, but changes
in time, approaching $1/H$ at late times, where $H$
is the phenomenological Hubble parameter.  Just as in
the $k=0$ FRW picture, the de Sitter model leads to
a rate of inflation that can vary dramatically with time,
starting at the Planck scale, going away during the
radiation- and matter-dominated phases, and re-emerging
at the current epoch.

This disussion of cosmology in the computational universe is 
necessarily qualitative and preliminary: a full discussion will
have to await the detailed solution of equations (2.3) in
candidate computational cosmologies, e.g., universes
in which the local computation corresponds to the standard
model.  We present this preliminary
picture based on the coarse-grained FRW equations simply to show
that the computational theory is apparently consistent with
observation for underlying homogeneous computations.   Though
preliminary, this discussion is nonetheless revealing: it suggests
both the existence of exotic, negative pressure matter that might
be identified with cold dark matter, together with a natural
mechanism for the production of dark energy.  In addition, the
current re-inflation of the universe by dark energy can take
place by the same mechanism as primordial inflation, taking into account 
the renormalization of the fundamental length scale given by 
the quantum geometric limit of the previous section.

\subsection{Matter in the computational universe}

Because quantum computations can support any local Hamiltonian
dynamics, there are many types of matter that can be supported
in the quantum computational model of quantum gravity.  
For example, the quantum computation could reproduce the
standard model as a lattice gauge theory, which would in
turn give rise to quantum gravity by `constructing' the 
distances between lattice points as described in section (2)
above.  

Any local quantum theory involving pairwise interactions
allows the construction of a theory of quantum gravity.
This suggests that we should search for quantum computations
whose local symmetries can reproduce the standard model.
In fact, the `swap' picture of quantum computation of section 1.2 above
possesses a local $SU(3) \times SU(2) \times U(1)$ symmetry:
$SU(3) \times U(1)$ commutes with the action of the gates, and
$SU(2)$ commutes with the action of the wires.  This symmetry
can be promoted to a gauge symmetry by looking at a version
of quantum computation that has $SU(3) \times U(1)$ transformations
at the two-qubit gates, and $SU(2)$ transformations on the
wires.  The swap picture of quantum computation and
the $SU(3) \times SU(2) \times U(1)$ version are simple
and standard pictures of quantum computation.  Whether or
not they can give rise to the standard model in a simple and
straightforward way remains to be seen.

Because the fundamental length scales depend on the cosmological
epoch in the computational universe, 
the masses of some of the elementary particles in the
theory are related to cosmological parameters such as 
the Hubble parameter.  In the quantum cellular
automaton models of the Dirac equation of Feynman {\it et. al.}
[62-65], the mass of the fundamental Dirac particle is
related to the amount of time it takes to accumulate
phases in the underlying quantum computation.
In particular, in the previous section it was shown both
for flat and for closed spatial computational cosmologies,
the amount of time it takes to accumulate
a phase of $\pi/2$ is $ \delta t = (4\pi/3)^{1/4} \sqrt{t_P/H}$.
This leads to a mass for the fundamental Dirac particle
of $ m c^2 = \pi \hbar/2\delta t  \approx 10^{-2}$ eV for 
the current observed values of $H$.  This value 
is consistent with observed constraints on the neutrino mass.
Again, whether or not this is a coincidence will have to
await a detailed calculation with exact microscopic models.

\subsection{Decoherent histories and the 
emergence of classical spacetime}

In the computational universe,
the emergence of the classical world can be described using the decoherent
histories approach [53-61] to quantum cosmology 
proposed by Gell-Mann and
Hartle [55].  In this approach, one assigns amplitudes to histories
of hydrodynamic variables such as local energy density, momentum
density, etc.  A natural application of decoherent histories to
the computational universe picture is to take the analog
of hydrodynamic variables to be averages $\bar T_{ab}$ of the 
components of $T_{ab}$ taken over coarse-grained volumes of space-time.  
Gell-Mann and Hartle [55] and Halliwell [56] have shown 
that in many situations,
such coarse-grained histories naturally decohere and give rise
to classical probabilities for the behavior of the coarse-grained
energy-momentum tensor.

Classical spacetime emerges in the computational universe by
the combination of the detailed quantum cosmological dynamics
discussed above, together with the the process of decoherence.  As shown, the
computational dynamics inflates the computation, spreading 
out the gates and making the discrete spacetime approximately
uniform.  In addition, as the computation evolves, different
computational histories begin to decohere from each other, giving
rise to a semiclassical structure to spacetime. 

In the decoherent histories approach [53-61], one constructs a decoherence
functional $D(H,H')$ where $H$ and $H'$ are coarse-grained histories
of the `hydrodynamic' variables $\bar T_{ab}$.  Hydrodynamic
variables arise naturally in the computational universe: indeed,
the program for deriving geometry from computation gives rise
exactly to a local definition of the energy-momentum tensor $T_{ab}$,
which yields $\bar T_{ab}$ upon coarse graining.
  
Each coarse-grained history consists of some `bundle' $C(H)$ of
underlying fine-grained computational histories.
Each fine-grained computational history within the bundle 
possesses a definite causal structure, definite local phases,
and a uniquely defined $T_{ab}$ and $R_{ab}$ that obey the
Einstein-Regge equations.
The decoherence functional in the path integral formulation [55] can
then defined to be
\begin{eqnarray}
D(H,H') = \sum_{C\in C(H), C' \in C(H')}
{\cal A}(C) \bar{\cal A}(C'),
\end{eqnarray}
where ${\cal A}(C)$ is the amplitude for the computational
history $C$.  Two coarse-grained histories are approximately decoherent if
$D(H,H') \approx 0$ for $H\neq H'$.
Decoherent histories of hydrodynamics variables behave effectively
classically.  The visible universe that we see around
us presumably corresponds to one such decoherent history.

As an example of the sort of calculation that is possible using
the decoherent histories approach, we follow [54-56] to show that coarse-grained
histories corresponding to different semi-classical trajectories 
for spacetime approximately decohere.    
Semiclassical histories for the energy-momentum tensor
correspond to histories which extremize the action.
In other words, such semiclassical histories correspond
to {\it stable} computations,  in which small changes in the routing of 
signals through the computation and in the angles rotated at each quantum logic
gate do not affect the overall amplitude of the computation to
first order.  In other words, stable computations are computational
analogs of histories that extremize the action. 

Such stable computations have a high amplitude by
the usual argument: computational histories in the vicinity
of a stable history have approximately the same phase, and
so positively interfere, leading to a relatively high probability
for stable computations.  In addition, stability also makes these
histories approximately decohere [54-56]. 

In particular, let $C(H)$ be the set of fine-grained
histories compatible with a particular coarse-grained semiclassical 
computational history $H$.  Let $C(H')$ be the set of fine-grained
histories compatible with a different coarse-grained semiclassical history
$H'$ within the same computation.  
As $C$ ranges over the fine-grained histories in $C(H)$, 
the phase of ${\cal A}(C)$ oscillates; the phase of ${\cal A}(C')$
oscillates independently.  As a result, performing the average over
fine-grained histories in equation (A3.9) yields
$D(H,H') \approx 0$ for $H,H'$ distinct coarse-grained histories
corresponding to different causal structures. At the same time, the
stability of the histories in $H$ gives high values of $D(H,H)$,
$D(H',H')$, thereby satisfying the condition for approximate 
decoherence [55-56], 
$|D(H,H')|^2 << D(H,H)D(H',H')$.  That is, coarse-grained
histories corresponding to distinct semiclassical histories tend to decohere.

Note that the degree of decoherence for coarse-grained histories
depends on the scale of the coarse graining.
Two completely fine-grained histories $H=C, H'=C'$ do not decohere,
as the decoherence functional $D(H,H')$ is then just equal
to the product of their amplitudes ${\cal A}(C) \bar{{\cal A}}(C')$.
So some coarse graining is required to get decoherence.

In other words, the computational universe naturally gives rise to 
a semiclassical spacetime via the decoherent histories approach to quantum
mechanics.  The computational universe picks out a special set of
fine-grained computational histories, corresponding to sequences of
projections onto the energy eigenspaces of the quantum logic gates.
Each fine-grained computational history gives rise to a spacetime
that obeys the Einstein-Regge equations.  Coarse graining 
yields hydrodynamic variables that are coarse-grained
averages of the energy-momentum tensor $T_{ab}$.  Coarse-graining
about stable histories yields hydrodynamic variables that approximately
decohere and obey classical probabilities.

\section{Which computation?}

Every quantum computation corresponds to a family of metrics,
each of which obeys the Einstein-Regge equations.  But which
computation corresponds to the universe we see around us?
What is the `mother' computation?  We do not yet know.   
Candidate computations must be investigated
to determine the ones that are compatible with observation.

The ability of quantum computers to simulate
lattice gauge theories suggests computations that can reproduce
the standard model of elementary particles.  Such quantum computations
could correspond to quantum cellular automata, with a regular, repeating
arrangement of vertices and edges [62-65]   
(care must be taken to insure that Lorentz invariance is 
preserved [62,66]). 
The local logic operations in a quantum cellular automaton can
be chosen to respect any desired local gauge symmetry such as a 
Yang-Mills theory.  Indeed, the results derived here show how
any underlying discrete quantum theory that gives rise
the standard model can be extended to a theory that gives
rise not only to the standard model, but to quantum gravity
coupled to the standard model as well.

Homogeneity and isotropy for a computational architecture can
also be enforced by introducing an element of randomness into the
wiring diagram.  For example, one can use a random computational
graph, as in the theory of causal sets [22-26]; 
the random arrangement of vertices 
insures approximate homogeneity and isotropy, in which a coarse-graining
containing $n$ vertices per cell is homogeneous and isotropic
to $O(1/\sqrt n)$.

An appealing choice of quantum computation is one which consists of
a coherent superposition of all possible quantum computations, as
in the case of a quantum Turing machine whose input tape is in
a uniform superposition of all possible programs (Methods,
section A4).  Such  
a `sum over computations' encompasses both regular and random
architectures within its superposition, and weighs computations
according to the length of the program to which they correspond:
algorithmically simple computations that arise from short programs
have higher weight.   The observational consequences of this
and other candidate computations will be the subject of future work.

\section{Future work}

This paper proposed a theory of quantum gravity derived from quantum
computation.  I showed that quantum computations naturally give
rise to spacetimes that obey the Einstein-Regge equations, with
fluctuations in the geometry of the spacetime arising from quantum
fluctuations in the causal structure and local action of the computation.
This theory makes concrete predictions for a variety of features
of quantum gravity, including the form of the back reaction of
metric to quantum fluctuations of matter, the existence of smallest
length scales, black hole evaporation, and quantum cosmology.
Some of these predictions, such as the absence of spontaneous
gravitationally induced decoherence, might be tested soon in the
laboratory [42].
In future work, these predictions will be explored in greater detail
using numerical simulations to calculate, for example, the spectrum
of curvature fluctuations in the early universe.

Because of the flexibility of quantum computation (almost any local
quantum system is capable of universal quantum computation), it is a
straightforward matter to give models of quantum computation that
exhibit a local Yang-Mills gauge invariance.  For example,
the `swap' picture of quantum computation of section 1.2 above
possesses a local $SU(3) \times SU(2) \times U(1)$ symmetry:
$SU(3) \times U(1)$ commutes with the action of the gates, and 
$SU(2)$ commutes with the action of the wires.  Whether or not this symmetry
can be identified with the $SU(3) \times SU(2) \times U(1)$ gauge symmetry 
of the Standard Model is a subject for future work. 

\vfill\noindent{\it Acknowledgements:}  This work was supported
by ARDA/ARO, AFOSR, DARPA, NSF, NEC, RIKEN, and the Cambridge-MIT Initiative.
The author would like to acknowledge useful conversations with 
O. Dreyer, F. Dowker, E. Farhi, M. Gell-Mann, 
V. Giovannetti, J. Goldstone, D. Gottesman,
A. Guth, J.J. Halliwell, A. Hosoya, L. Maccone, 
D. Meyer, F. Marcopoulou, J. Oppenheim, P. Zanardi, and many others.
\eject

\noindent{\bf References:}

\bigskip\noindent 
[1] M.A. Nielsen, I.L. Chuang, {\it Quantum Computation and Quantum
Information}, Cambridge University Press, Cambridge, 2000. 

\bigskip\noindent[2]
R.P. Feynman, {\it Int. J. Theor. Phys.} {\bf 21}, 467 (1982).

\bigskip\noindent
[3] S. Lloyd, {\it Science}, {\bf 273}, 1073-1078 (1996).

\bigskip\noindent[4] D.S. Abrams, S. Lloyd, {\it Phys. Rev. Lett.} {\bf 79},
2586-2589 (1997).

\bigskip\noindent[5] S. Lloyd, `Unconventional Quantum Computing Devices,'
in {\it Unconventional Models of Computation}, C. Calude, J. Casti,
M.J. Dinneen eds., Springer (1998).

\bigskip\noindent[6] S. Bravyi., A. Kitaev, {\it Ann. Phys.} {\bf
298}, 210-226 (2002); quant-ph/0003137.

\bigskip\noindent [7]
C. Rovelli, {\it Quantum Gravity}, Cambridge University Press,
Cambridge, 2004.

\bigskip\noindent [8]
S. Carlip, {\it Rept. Prog. Phys.} {\bf 64}, 885 (2001); gr-qc/0108040.

\bigskip\noindent [9]
A. Ashtekar, {\it Lectures on Non-perturbative Canonical Gravity,}
(Notes prepared in collaboration with R.S. Tate),
World Scientific, Singapore, 1991.

\bigskip\noindent [10]
G.W. Gibbons, S.W. Hawking, {\it Euclidean Quantum Gravity},
World Scientific, Singapore, 1993.

\bigskip\noindent [11]
G. 't Hooft, {\it Class. Quant. Grav.} {\bf 16} 3263-3279 (1999),
gr-qc/9903084; {\it Nucl. Phys. B} {\bf 342}, 471 (1990).

\bigskip\noindent [12]
J. Wheeler, in {\it Proceedings of the 3rd International 
Symposium on Foundations of Quantum Mechanics}, Tokyo (1989);
{\it Geons, Black Holes, \& Quantum Foam: A Life in Physics},
by John Archibald Wheeler with Kenneth Ford, W.W. Norton, (1998).

\bigskip\noindent [13]
P.A. Zizzi, {\it Gen. Rel. Grav.} {\bf 33}, 1305-1318 (2001),
gr-qc/0008049; gr-qc/0409069.
 
\bigskip\noindent [14]
J. Polchinski, {\it String
Theory}, Cambridge University Press, Cambridge (1998).

\bigskip\noindent [15]
C. Rovelli,  L. Smolin, {\it Nucl. Phys. B} {\bf 331}, 80-152 (1990).

\bigskip\noindent [16]
C.Rovelli, {\it Phys. Rev. D} {\bf 42}, 2638 (1990);
{\it Phys. Rev. D} {\bf 43}, 442 (1991);
{\it Class. Quant. Grav.} {\bf 8}, 297 (1991);
{\it Class. Quant. Grav.} {\bf 8}, 317 (1991).

\bigskip\noindent [17]
C. Rovelli, L. Smolin, {\it Phys. Rev. D} {\bf 52}, 5743-5759 (1995).

\bigskip\noindent [18]
J.C. Baez, {\it Adv. Math.} {\bf 117}, 253-272 (1996).

\bigskip\noindent [19]
J.W. Barrett, L. Crane, {\it J. Math. Phys.} {\bf 39}, 3296-3302 (1998).

\bigskip\noindent [20]
J.C. Baez, J. D. Christensen, T.R. Halford, D.C. Tsang
{\it Class. Quant. Grav.} {\bf 19}, 4627-4648 (2002); gr-qc/0202017.

\bigskip\noindent [21] A.D. Sakharov, {\it Doklady Akad. Nauk S.S.S.R.}
{\bf 177}, 70-71 (1967); English translation {\it Sov. Phys. Doklady}
{\bf 12}, 1040-1041 (1968).

\bigskip\noindent [22]
L. Bombelli, J. Lee, D. Meyer, R.D. Sorkin, {\it Phys. Rev. Lett.}
{\bf 59}, 521-524 (1987).

\bigskip\noindent [23]
X. Matin, D. O'Connor, D. Rideout, R.D. Sorkin, {\it Phys. Rev. D.}
{\bf 63}, 084026 (2001).

\bigskip\noindent [24]
G. Brightwell, H.F. Dowker, R.S. Garcia, J. Henson, R.D. Sorkin
{\it Phys. Rev. D} {\bf 67}, 08403 (2003); gr-qc/0210061.

\bigskip\noindent [25]
E. Hawkins, F. Markopoulou, H. Sahlmann,
{\it Class. Quant. Grav.} {\bf 20} 3839 (2003);
hep-th/0302111.

\bigskip\noindent [26]
D. Dou, R.D. Sorkin, {\it Found. Phys.} {\bf 33}, 279-296  (2003);
 gr-qc/0302009.

\bigskip\noindent [27]
A. Kempf, {\it Phys. Rev. Lett.} {\bf 85}, 2873 (2000),  hep-th/9905114;
{\it Phys. Rev. Lett.} {\bf 92}, 221301 (2004), gr-qc/0310035.

\bigskip\noindent [28]
D. Finkelstein, {\it Int. J. Th. Phys.} {\bf 27}, 473 (1988).

\bigskip\noindent [29]
A. Marzuoli, M. Rasetti, {\it Phys. Lett. A}{\bf 306},
79-87 (2002), quant-ph/0209016; quant-ph/0407119; quant-ph/0410105.

\bigskip\noindent [30]
F. Girelli, E.R. Livine, gr-qc/0501075.

\bigskip\noindent [31] M. Levin, X.-G. Wen, {\it Rev. Mod. Phys.}
{\bf 77}, 871-879 (2005), cond-mat/0407140.

\bigskip\noindent [32]  D. Bacon, J. Kempe, D.A. Lidar, 
K.B. Whaley, {\it Phys. Rev. Lett.} {\bf 85}, 1758-61 (2000),
quant-ph/9909058;
{\it Phys. Rev. A} {\bf 63}, 042307 (2001), quant-ph/000406.

\bigskip\noindent [33]
D. P. DiVincenzo, D. Bacon, J. Kempe, G. Burkard, K. B. Whaley,
{\it Nature} {\bf 408}, 339-342 (2000).

\bigskip\noindent [34]
A. Einstein, {\it Ann. Physik} {\bf 49} (1916), in
H.A. Lorentz, A. Einstein, H. Minkowski, H. Weyl,
{\it The Principle of Relativity}, Dover, New York (1952).

\bigskip\noindent [35]
T. Regge, {\it Nuovo Cimento} {\bf 19}, 558-571 (1961).

\bigskip\noindent [36]
W.A. Miller, {\it Class. Quant. Grav.} {\bf 14} L199-L204 (1997); gr-qc/9708011.

\bigskip\noindent [37]
R. Friedberg, T.D. Lee, {\it Nucl. Phys. B} {\bf 242}, 145-166 (1984).

\bigskip\noindent [38] A. Okabe, B. Boots, K. Sugihara, {\it Spatial
Tessellations: Concepts and Applications of Voronoi Diagrams},
Wiley, Chichester (1992).


\bigskip\noindent [39]
S.W. Hawking, G.F.R. Ellis, {\it The Large Scale Structure of
Space-Time}, Cambridge University Press, Cambridge, 1973.

\bigskip\noindent [40] J.B. Hartle, R. Sorkin, {\it Gen. Rel.
Grav.} {\bf 13}, 541-549 (1981).

\bigskip\noindent [41]
J. Ambjorn, J. Jurkiewicz, R. Loll, {\it Phys. Rev. Lett.} {\bf 93,}
131301 (2004); hep-th/0404156.

\bigskip\noindent [42]
W. Marshall, C. Simon, R. Penrose, D. Bouwmeester,
{\it Phys. Rev. Lett.} {\bf 91}, 130401 (2003);
quant-ph/0210001.

\bigskip\noindent [43] 
S. Lloyd, ``Almost certain escape from black holes,'' quant-ph/0406205.
To appear in {\it Phys. Rev. Lett.}

\bigskip\noindent [44]
G.T. Horowitz and J. Maldacena, ``The Black-Hole Final State,''
hep-th/0310281. 

\bigskip
\noindent [45] D. Gottesman and J. Preskill, ``Comment on `The black hole final
state,' " hep-th/0311269.

\bigskip
\noindent [46] U. Yurtsever and G. Hockney, ``Causality, entanglement,
and quantum evolution beyond Cauchy horizons,'' quant-ph/0312160.

\bigskip
\noindent [47] U. Yurtsever and G. Hockney,
``Signaling and the Black Hole Final State," quant-ph/0402060.

\bigskip\noindent [48]
V. Giovannetti, S. Lloyd, L. Maccone,
{\it Science} {\bf 306}, 1330 (2004),
quant-ph/0412078; quant-ph/0505064. 

\bigskip\noindent [49] J.D. Bekenstein, {\it Phys. Rev. D} {\bf 7}, 2333 (1973);
{\it Phys. Rev. D.} {\bf 23}, 287 (1981);
{\it Phys. Rev. Letters\/} {\bf 46}, 623 (1981);  {\it Phys. Rev. D.}
{\bf 30}, 1669-1679 (1984).

\bigskip\noindent [50]
G. 't Hooft, ``Dimensional reduction in quantum gravity,'' gr-qc/9310026.

\bigskip\noindent [51] 
L. Susskind, {\it J. Math. Phys.} {\bf 36},
6377, hep-th/9409089.

\bigskip\noindent [52]
Linde, A., {\it Particle Physics and Inflationary Cosmology},
Harwood Academic, New York (1990).

\bigskip\noindent [53]
R. Griffiths, {\it J. Stat. Phys.} {\bf 36}, 219 (1984).

\bigskip\noindent [54]
R. Omn\'es, {\it J. Stat. Phys.} {\bf 53}, 893, 933, 957 (1988); 
 {\it J. Stat. Phys.} {\bf 57}, 359 (1989); {\it Rev. Mod. Phys.}
{\bf 64}, 339 (1992); {\it The Interpretation of Quantum
Mechanics}, Princeton University Press, Princeton, 1994.

\bigskip\noindent [55]
M. Gell-Mann, J.B. Hartle
{\it Phys. Rev. D} {\bf 47}, 3345-3382 (1993); gr-qc/9210010.

\bigskip\noindent [56] J.J. Halliwell, {\it Phys. Rev. D} 
{\bf 58} 105015 (1998), quant-ph/9805062; 
{\it Phys. Rev. D} {\bf 60} 105031 (1999), quant-ph/9902008;
{\it Phys. Rev. Lett.} {\bf 83} 2481 (1999), quant-ph/9905094; 
`Decoherent Histories for Spacetime Domains,' in 
{\it Time in Quantum Mechanics,} edited by J.G.Muga, R. Sala Mayato 
and I.L.Egususquiza (Springer, Berlin, 2001), quant-ph/0101099.

\bigskip
\noindent [57]
J.J. Halliwell, J.Thorwart, {\it Phys. Rev. D} {\bf 65} 104009 (2002);
gr-qc/0201070.

\bigskip\noindent [58] J.B. Hartle,  {\it Phys. Scripta T}{\bf 76}, 67 (1998),
gr-qc/9712001.

\bigskip\noindent [59] F. Dowker and A. Kent, {\it Phys. Rev. Lett.} {\bf 75},
3038 (1995).

\bigskip\noindent [60] A. Kent, {\it Physica Scripta} {\bf T76}, 78 (1998).

\bigskip\noindent [61] T.A. Brun, J.B. Hartle, {\it Phys. Rev. E}{\bf 59},
6370-6380 (1999), quant-ph/9808024;
{\it Phys. Rev. D}{\bf 60}, 123503 (1999), quant-ph/9905079.

\bigskip\noindent [62]
I. Bialynicki-Birula,  {\it Phys. Rev. D} {\bf 49}, 6920-6927 (1994);
hep-th/9304070. 

\bigskip\noindent [63]
Margolus, N., {\it Ann. N.Y. Acad. Sci.} {\bf
480}, 487-497 (1986); and in
{\it Complexity, Entropy, and the Physics of Information, vol. VIII,}
W. H. Zurek, ed., pp. 273-87, Addison-Wesley, Englewood Cliffs 1990.

\bigskip\noindent [64]
R.P. Feynman, {\it Rev. Mod. Phys.} {\bf 20}, 267 (1948); R.P.
Feynman, A.R. Hibbs, {\it Quantum Mechanics and Path Integrals},
McGraw-Hill, New York, 1965.

\bigskip\noindent [65]
B. Gaveau, T. Jacobson, M. Kac, L.S. Schulman, {\it Phys. Rev. Lett.}
{\bf 53}, 419 (1984).

\bigskip\noindent [66]
F. Dowker, J. Henson, R.D. Sorkin, {\it Mod. Phys. Lett. A}{\bf 19} 
1829-1840 (2004); gr-qc/0311055.
     
\section{Supporting online material: Methods}

\subsection{Introduction} 

Note that while the computational universe program is not
obviously related to string theory and loop quantum gravity,
it is not necessarily incompatible with these approaches.  
(For work relating quantum computation to loop quantum
gravity see [13, 29-30].)
First of all, to the extent that the dynamics
of these theories are discrete and local, they can be
efficiently reproduced on a quantum computer.  So, for example,
a quantum computer can reproduce the dynamics of discretized
conformal field theories, which should allow the simulation
of string theory in the anti-de Sitter/conformal field theory
(AdS/CFT) correspondence.  In addition, 
it is likely that string theory and loop quantum gravity are 
themselves computationally universal in the sense that they can 
enact any quantum computation.  (Computational universality
is ubiquitous: most non-trivial dynamics are
computationally universal.)  If this is so, it is possible
that all of these approaches to quantum gravity are logically equivalent in 
the same sense that a Macintosh is logically equivalent to
a PC which in turn is logically equivalent to a Turing machine.

Such logical equivalence need not necessarily imply {\it physical} 
equivalence, however.  Computation has consequences: it induces
spacetime curvature.  In the model of quantum gravity presented here,
the extra computation required to reproduce string theory from
an underlying computation or from loop quantum gravity might
in principle be detectable.  This issue will be discussed
further below.

The idea of deriving gravity from an underlying quantum theory
is reminiscent of Sakharov's work deriving
gravity from an underlying elastic medium [21].  Unlike 
[21], however, the theory presented here presents an
explicit mechanism for the back reaction of gravity to quantum
fluctuations in the underlying matter.

\subsection{The computational graph}

As just noted, there are many different models of universal quantum
computation, all logically equivalent to each other in the sense
that each model can simulate the others efficiently (it is even possible to 
have quantum circuit models that admit closed cycles) [1].   
The quantum circuit model for quantum computation adopted here is 
the model most closely related to the structure of events in spacetime.  

Many models of quantum computation use single-qubit operations.
For example, a popular model uses single-qubit rotations together
with controlled-NOT operations or swap operations [1].  
Single-qubit rotations together with swap operations possess
an $SU(2)$ gauge symmetry which makes this model richer and
more complex than the basic two-qubit logic gate model
discussed here.  Quantum computation with gauge symmetries
is related to computational models of elementary particles
and will be discussed elsewhere.  The addition of single-qubit
logic gates will be discussed in section A2.3 below.

Note that for a quantum computer to be able to simulate fermionic
systems efficiently, it should have access to fermionic degrees
of freedom [2-6].    This is consistent with the identification of qubits
with spin-1/2 massless particles.  Fermionic statistics can
be enforced either by defining quantum logic gates in terms
of anti-commuting creation and annihilation operators [5], or
by using appropriate local interactions [6,31] which enforce fermionic
statistics dynamically. 
  
\subsection{Computational histories}

In the example given in section (1.2), for simplicity we took the
lowest eigenvalue to be zero and the only non-zero eigenvalue
to be 1.  For the generic case in which quantum logic gates have Hamiltonians
with more than one non-zero eigenvalue, there is more than one kind of 
scattering event.  Because only the relative phase is observable,
we take the lowest eigenvalue to be zero and
the other eigenvalues to be positive.  
The zero eigenvalue still corresponds to
non-scattering non-events; but now when scattering occurs, the qubits
can acquire different phases, one for each non-zero eigenvalue.
The number of causal structures is the same as discussed in section (1.2)
(i.e., $2^n$, where $n$ is the number of gates), but now each causal
structure gives rise to a number of different computational histories
at each gate, each with a different action corresponding to the
different phase for each non-zero eigenvalue.  If there are
$m$ distinct eigenvalues per gate, then there are $m^n$ computational
histories.   

Taking the eigenvalues of the Hamiltonian and phases
to be non-negative insures the positivity of the energy of
interaction in quantum logic gates as measured locally.  
More precisely, as shown in section (A2.2) below,
it assures the negativity of ${T^a}_a$ and the positivity of
curvature.   That is, the local non-negativity of the energy of
interaction within a quantum logic gate insures a version of the 
strong energy condition ([39] section 4.3).

If acquired phases and local energies can be negative, then
the procedure of assigning a metric to spacetime based on
the underlying quantum computation still holds.  Now, however,
it is possible to fit arbitrarily large amounts of
computation into a volume of spacetime without generating
any curvature on average, merely by having the average angle
rotated within the volume be zero.  The inclusion of negative
local energies and phases has implications for holography and
quantum cosmology and will be discussed further in section 3
below.

The relationship between the overall computational
graph and that of the individual computational histories works
as follows.  Each logic gate has two inputs,
$A$ and $B$, and two outputs, $A'$ and $B'$.  The inputs
$A$ and $B$ come from outputs
$A_{\rm out}$ and $B_{\rm out}$ of other gates, and the outputs
$A'$ and $B'$ are inputs
to $A_{\rm in}$ and $B_{\rm in}$ of other gates.  In the case that there is
scattering, there is a vertex at the gate.  In the case of
no scattering, delete the vertex and connect $A_{\rm out}$ directly
to $A_{\rm in}$, and $B_{\rm out}$ directly to $B_{\rm in}$.  
The new connections are null lines in the computational history where there is
no scattering at that gate.
        
The process of excising a gate to construct a new computational
history can be thought of as follows: the gate
is cut out, the cut lines are re-connected to the appropriate
ends, and then they `spring tight' to give null lines in the
new computational history from which the non-scattering vertex
has been deleted.

\subsection{Constructing a Simplicial Lattice}

A computational history does not form a simplicial lattice in
itself, but it can be extended to a simplicial lattice by
adding additional edges, and, if necessary, vertices.  There
are a variety of ways to extend a graph to a simplicial lattice
[38].  Here we discuss two, but more are possible.  Both
constructions are based on dual Delaunay and Voronoi lattices.
Method 1 results in a discretized manifold that is topologically
equivalent to $R^4$, but requires the addition not only of more
edges, but of more vertices as well.  Method 2 does not require
additional vertices to be added, but typically results in a 
topologically non-trivial discretized manifold.  The important
feature of the different possible triangulations is that
all possible ways of extending the computational history into a
simplicial lattice, $D_C$, result in spacetimes that are equivalent
in terms of their observable content, i.e., their causal structure,
action, and curvature.  Accordingly, one may adopt whichever
triangulation is convenient for a particular problem.

\bigskip\noindent{\it Method 1}

\smallskip\noindent
(1) Embed the computational history in $R^4$ via
an embedding mapping $\chi(C)$ that maps vertices to points in $R^4$ 
and wires to line segments in the manifold connecting the vertices, as above.

\smallskip\noindent
(2) Construct the Voronoi lattice $V_C$ for the vertices of
$ \chi(C)$ by dividing spacetime into regions that are
closer to one vertex of $C$ than to any other vertex.
Here, `closer' is defined using the ordinary Euclidean
metric on the manifold.
The Voronoi lattice defines the volumes associated
with vertices of the computational graph.

\smallskip\noindent
(3) Connect vertices of the computational graph to neighboring
vertices through the centroids of faces of the Voronoi lattice
to form the Delaunay lattice $D_C$ dual to $V_C$.  $D_C$ is
the simplicial lattice that will be used to define the Regge calculus.
(In some degenerate cases of measure 0, $D_C$ is not simplicial; in these
cases a slight change in the embedding will render $D_C$ simplicial.)
The faces of the Voronoi lattice bisect the edges of the Delaunay lattice.

\smallskip\noindent
(4) Check to make sure that the original edges of the computational
history are edges of $D_C$.  If they are not, simply add new,
`reference' vertices along the edges of $C$ and repeat steps $1-3$.
By making the reference vertices sufficiently close together,
this procedure insures that all edges of $C$ fall along edges
of $D_C$.  In particular, if the reference vertices along a given
are closer to their neighbors along the edge than to any other vertices,
then every edge of $C$ is guaranteed to lie along edges of $D_C$    
The reference vertices will in general divide the original edges of $C$ 
into several segments.

Steps (1-4) are a standard procedure for constructing a simplicial
Delaunay lattice together with its dual Voronoi lattice.

Some authors exhibit a prejudice against the use of highly
elongated simplices in a Delaunay lattice [38].
If desired, further reference vertices can be added in `empty space'
to construct a Delaunay lattice without elongated simplices.
Once again, just as different surveyors can choose different
triangulations for the same landscape,
the precise arrangement of added edges and vertices 
does not affect the observable content of the
spacetime.  These additional edges and vertices are `fictions'
that we use to fill in a simplicial geometry: the lengths of
these added edges are derived from the underlying observational
content.   In contrast, the `factual' content of the geometry lies
in its causal structure and local actions.

Note that if we choose to add additional reference vertices, we 
must give a method for assigning the as yet unknown terms of the
metric at those vertices.  That method is closely analogous to
the technique for assigning the unknown terms of the metric at
the vertices of the computational history and will be given in
section A2.3 below.

\bigskip\noindent{\it Method 2}

Method 1 for constructing a simplicial lattice results
in a discretized manifold that is topologically equivalent to $R^4$ or to
a convex subset of $R^4$. 
A second way of extending the computational history to a simplicial
lattice requires no additional reference vertices, but can result
in a discretized manifold that is not simply connected.  Once again, 
the resulting lattice gives a spacetime with the same causal structure, action,
and curvature as method 1, i.e., any difference in the topology of
the different lattices is observationally undetectable.  

In method 2, to construct a simplicial lattice corresponding to 
the discrete computational history, consider not just one embedding
map $\chi$, but a sequence of embedding maps $\{ \chi_\ell \}$.  
Each embedding map $\{ \chi_\ell \}$ embeds a vertex $\ell$ in $R^4$
together with a neighborhood $\nu_\ell$ of nearby vertices, including
its neighbors and next-to-nearest neighbors.  $\chi_\ell$ is one
chart in the atlas of embedding maps $\{\chi_\ell\}$ that defines
the discretized manifold.  Now, once again, use the Delaunay/Voronoi
construction to produce a simplicial lattice for the vertices in
the neighborhood.   Because there are only a small number of vertices
in the neighborhood, it is always possible to choose the embedding
so that each edge in the neighborhood is an edge in the Delaunay
lattice.  Similarly, in the embeddings can always be chosen
to give the same set of Delaunay edges and Voronoi volumes
in the overlap $\nu_\ell \cap 
\nu_{\ell'}$ between two adjacent vertices $\ell$ and $\ell'$.
The resulting atlas of overlapping charts, together with the
accompanying Delaunay and Voronoi lattices, defines the discretized
manifold $M_C$.  Note that in this construction no new reference
vertices are required.

The price of adding no reference vertices is that the resulting
discretized manifold may no longer be simply connected.
In method 2, simple, locally-connected computations such as
quantum cellular automata typically give rise to low genus
$M_C$, while more complex computations with long-distance connections
can give rise to manifolds with high genus.  

No matter how the computational history is extended to a simplicial
lattice, the resulting discrete spacetime possesses the same
causal structure and action.  As will now be seen, it also
possesses the same curvature at each vertex.

\subsection{Regge Calculus}

The four null lines at each vertex of the computational graph
set four out of the ten components
of the metric in four dimensions (i.e., in the null basis given
by the lines, the on-diagonal parts of the metric are zero).
The four null lines at each vertex define six planes, one for
each pair of lines and for each of the undetermined off-diagonal
components of the metric.  Two of these
planes -- the plane defined by the two ingoing null lines, and the
one defined by the two outgoing null lines -- contain vectors that
are timelike with respect to the vertex.  The remaining four -- 
defined by one ingoing and one outgoing line -- contain spacelike
vectors.  Each plane corresponds to an off-diagonal term in the
metric. 

Because quantum logic gates have two inputs and two outputs,
four dimensions arise naturally in the computational universe.
The local null tetrad over-determines the metric in fewer
than four dimensions, and under-determines it in more than
four dimensions.  In the computational universe, the four-dimensional
structure of spacetime arises out of pairwise interactions between
information-bearing degrees of freedom.  Note that as in [41]
one must still verify that a microscopically four-dimensional
spacetime still looks four-dimensional at macroscopic scales.

\subsection{The Einstein-Regge equations}

To obtain the explicit form of the Einstein-Regge equations,
$\delta I_G/ \delta g_{ab}(\ell) +
\delta I_C/ \delta g_{ab}(\ell) = 0$,
recall that the gravitational action is
$I_G = (1/8\pi G) \sum_h \epsilon_h A_h$,
where $\epsilon_h$ is the deficit angle of the hinge $h$ and $A_h$ is
its area.   Under a variation of the metric at point $\ell$,
$g_{ab}(\ell) \rightarrow g_{ab}(\ell) + \delta g_{ab}(\ell)$,
the variation in the gravitational action is given by Regge [12]:
\begin{eqnarray}
\delta I_G = {1\over 8\pi G} \sum_h \epsilon_h \delta A_h \\
= {1\over 16\pi G} \sum_h \sum_{p(h)} l_{ph} \delta l_{ph} {\rm cotan}
\phi_{ph}. 
\end{eqnarray}
Here, as Regge showed, the variation in $\epsilon_h$ cancels out,
$l_{ph}$ is the $p$'th edge of hinge $h$ (p=1,2,3), and
$\phi_{ph}$ is the angle in the hinge opposite to $l_{ph}$.

Explicitly inserting the variation in the metric, we obtain
\begin{eqnarray}
\delta I_G = {1\over 8\pi G} \sum_h \epsilon_h {\delta A_h \over
\delta g_{ab}}  \delta g_{ab} \\
= {1\over 16\pi G} \sum_h \sum_{p(h)} l_{ph}
{\delta l_{ph} \over  \delta g_{ab}}  \delta g_{ab} {\rm cotan} \phi_{ph}.
\end{eqnarray}
Combining equation this equation with equation (2.2) for the variation
of the action of the computational matter with respect to the metric
gives
\begin{eqnarray}
-\sum_{h\in N(\ell)} \sum_{p(h)} l_{ph}
{\delta l_{ph} \over  \delta g_{ab}(\ell)}
{\rm cotan} \phi_{ph} =
8 \pi G  ( \hat T^{ab} - {\cal U}_{\ell} g^{ab}(\ell) )
\Delta V_\ell.
\end{eqnarray}
Here, the sum over hinges includes only hinges $h\in N(\ell)$
that adjoin the vertex $\ell$, as the edges of other hinges do not change
under the variation of the metric at $\ell$.
More succinctly, we have 
\begin{eqnarray}
-\sum_{h\in N(\ell)} \epsilon_h {\delta A_h \over \delta g_{ab}}
= 4\pi G T^{ab} \Delta V_\ell
\end{eqnarray}
This is the expression of the Einstein-Regge equations in 
the computational universe.

\subsection{Satisfying the Einstein-Regge equations}

In sections 2.2-2.3, all vertices in the simplicial lattice were
assumed to be vertices of the computational history, as in
method 2 above.  That is, each vertex has four null edges,
corresponding to two incoming and two outgoing quantum degrees
of freedom.  For method 1, we have to assign the unknown
components of the metric and of the energy-momentum tensor
at reference vertices where there is only one null line,
corresponding to a single edge of the computational history,
or no null lines, corresponding to a reference vertex put down
in `empty space.'  

For the case of a reference vertex with one null line $E_1$, simply
proceed as in section 2.2.  Using that null line as one vector
in a basis at that vertex, we see that only one on-diagonal
term in the metric is known, e.g., $g_{11} = 0$.  The remaining
on-diagonal terms and the off-diagonal terms are to be assigned
by the requirement of self consistency, as before.   
At a reference vertex, the action of the `computational matter'
is zero, and the only non-zero term in the energy momentum
tensor is the $11$ kinetic energy term:
$\hat T^{ab} = \gamma_1 E_1^a E_1^b$.  This term corresponds
to a single massless particle propagating in the $E_1$ direction. 
Here also, the value of the particle's kinetic energy density, $\gamma_1$,
is to be determined by self-consistency.
To satisfy the Einstein-Regge equations (2.3), first assign the
unknown terms of the metric at the vertex to make all but
the $11$ term in equation (2.3) hold.   The metric is now completely
determined. Next, assign the kinetic energy density $\gamma_1$ to
make the $11$ component of (2.3) hold.  The Einstein-Regge
equations are now satisfied at the reference vertex.

Initial and final states of the computation correspond to vertices
that have only one null line emerging from them (initial states)
or leading to them (final states).  If the initial and final
states are two-particle singlet states, then they have two
null lines emerging from them or leading to them.  The method
for satisfying the Einstein-Regge equations at such points
is essentially the same as for the reference vertices of
the previous paragraph.  The null line or lines at the vertex
determines part of the metric at that point; select the remaining
parts of the metric to insure that the Einstein-Regge
equations hold for the remaining parts of the metric
at that point.  (Here one must be sure to
include the boundary terms [40] in the action.)  Then
assign the kinetic energies for the null lines to make the
rest of the Einstein-Regge equations hold.  

The case of a reference vertex in `empty space' is even simpler.
Here, nothing is known about the metric {\it a priori}.  All the terms in
the energy-momentum tensor are zero.  Equation
(2.3) is then a ten-component equation in the ten unknown components
of the metric.  Solving equation (2.3) assigns those components
in a self-consistent fashion.  Essentially, in parts of spacetime
where there are no logic gates or quantum wires, i.e., no information 
and no information processing, we assign the metric by solving the
vacuum Regge equations.  In the absence of computational `matter,' 
the only part of
the curvature tensor that is non-zero is the Weyl tensor, corresponding
to gravity waves.  As usual in general relativity, the sources of
those gravity waves are places where there is matter,
and the Ricci tensor and curvature scalar are non-zero (see [39],
section 4.1, and section 3 below).  Note that gravity waves
are not quantized directly in this theory: instead, their presence
is deduced by their interactions with the computational matter.

\subsection{Single qubit quantum logic gates}

Single qubit quantum logic gates with one input and one
output can be accommodated in the
theory in a method similar to that used in the preceding
paragraphs.  A single-qubit gate gives rise to several
different computational histories, one corresponding to
each distinct eigenvalue of the gate's unitary operator.
For the zero eigenvalue (no action), there is no
vertex in the computational history associated with the
gate, and the input and output edges of the gate are
joined into a single edge, corresponding to a single
null line in the computational spacetime.  
For non-zero eigenvalues $\theta$, the gate gives rise to a vertex
with action $-\hbar \theta$ in the computational history.
The input and output lines then correspond to two linearly
independent null lines in the spacetime.  

These two lines define two of the components of the metric:
the on-diagonal terms corresponding to the two lines are zero.
Choose the remaining eight components of the metric to make
the corresponding eight components of the Einstein-Regge
equations (2.3) hold.  Finally, choose the kinetic energy
terms for the `particles' travelling along those two null
lines to make the remaining two components of the Einstein-Regge
equations hold.

\subsection{Comparison with GPS}

The procedure described in section 2.3 for inferring the off-diagonal
part of the metric and the traceless part of the energy-momentum tensor 
is analogous to classical procedures for determining the geometry
of spacetime from the behavior of matter within that spacetime [39]. 
The reconstruction of the metric from the null lines and the local
action at each vertex closely resembles the way that a swarm
of GPS satellites maps out spacetime geometry by sending signals
along null lines and timing their arrival by counting the ticks
of their local clocks [48].  In the method for reconstructing the metric
detailed here, a `tick' of a clock corresponds
to a local accumulation of an angle $\pi$ at a quantum logic gate. 

The method for reconstructing spacetime geometry given here is
slightly more complicated than the reconstruction of geometry
from GPS data, however, because we also have to determine the
form of the energy-momentum tensor.  In particular, the atomic
clocks in GPS satellites rely on the fact that all the atoms
in the clock have well-characterized hyperfine energy level splittings 
that are in principle identical from atom to atom. 
Here, we don't know the energy and time scales corresponding to
a given quantum logic gate until the length scale, as embodied
by the conformal factor, has been assigned at each gate.
Still, as shown above, exactly enough information is embodied in
the behavior of the computational `matter' -- where it goes, and what it does
when it gets there -- to deduce the full metric and the energy-momentum 
tensor.

\subsection{Coarse graining and the Einstein equations}

Every computational
history corresponds to a spacetime that obeys the Einstein-Regge
equations.  The overall quantum computation is the sum of
individual computational histories corresponding to different
causal structures.  Fluctuations in space and
time track the quantum fluctuations in the local routing of
information and action accumulated at each quantum logic gate.

Solving (2.3) to assign the off-diagonal part of the metric 
and the on-diagonal part of the energy-momentum tensor
involves solving a nonlinear partial difference equation, and
then determining the on-diagonal part of the energy-momentum
tensor from the remainder of the Einstein-Regge equations.
Future work will involve solving (2.3) directly in order
to investigate the behavior of the computational universe
at the smallest scales.  

The underlying dynamics in the computational universe is given
by Regge calculus.  Einstein's equations are known to follow
from Regge calculus by a procedure of coarse graining [35].
This coarse-grained picture proceeds as follows.
Coarse grain spacetime by averaging over a
cell size $\zeta$ where the
volume $dv \equiv \zeta^4 $ in the coarse graining contains many points
in the computational graph.  Then construct a coarse-grained
metric $g_{ab}(x)$, and coarse-grained curvatures $R(x)$
and Lagrangians ${\cal L}(x)$ by averaging
over those coarse-grained volumes.

As Regge noted (and as is true for all coarse grainings
of discrete equations) care must be taken to insure that
the coarse-grained equations accurately reflect the true 
behavior of the underlying discrete dynamics.  Such a
coarse grained picture should be reasonably accurate when
\smallskip\noindent
(1) The underlying computational graph is relatively
homogeneous, as it is for cellular automata and homogeneous
random architectures.
\smallskip\noindent
(2) The curvature is significantly smaller than the
inverse Planck length squared.
\smallskip\noindent
(3) The variation of the average angle rotated per coarse-grained
cell is small from cell to cell.  

\noindent Note that the quantum computations that reproduce
homogeneous, local quantum theories such as lattice gauge
theories obey all three of these requirements. 

Coarse graining is a useful tool for analyzing nonlinear discrete systems
such as those considered here.  But the realm of validity of
coarse graining for nonlinear systems is a subtle subject;
although heuristic arguments like those above can give
some confidence in the application of coarse graining,
in the final analysis its validity must be tested by 
comparing coarse-grained solutions with the complete solution
of the underlying discrete dynamics.  In the computational
universe, to analyze scenarios such as initial and final
singularities where quantum gravity truly comes into play,
the underlying equations must be solved directly.

With these caveats in mind,
now follow the procedure for deriving the Einstein-Regge
calculus above, but in the coarse-grained context.
In the coarse-grained setting, our equations take the conventional
Einstein form:
\begin{eqnarray}
R_{ab} - {1\over 2} g_{ab} R = 8\pi G(\hat T_{ab} - g_{ab}{\cal U})
= 8\pi G T_{ab}.  
\end{eqnarray}
We proceed as for the discrete case.  The underlying computational
graph gives us four null lines at each point, fixing the 
on-diagonal parts of the metric to zero in a coordinate
system for which those lines form a basis.  Assign the six unknown 
off-diagonal parts of the metric to make the off-diagonal part
of the Einstein equations hold.  The left-hand side of the Einstein
equations is now completely determined. 
Once the off-diagonal part of the metric has been obtained,
the on-diagonal kinetic energy terms $\hat T^{ab}$ can be obtained as before
by assigning them so that the on-diagonal part of Einstein's equations
are obeyed.  The full Einstein equations now hold.

To assign the metric uniquely so that
make Einstein's equations hold globally, boundary conditions must be
supplied for equation (A2.4).
At the initial and final points of the computation,
the metric is incompletely defined, corresponding to a singularity
in the equations.  This suggests setting boundary conditions
by taking $\Omega=0$ at these points.  In the case of an
infinite computation, with no final points, one can set
boundary conditions by taking
$\Omega =0$, $\Omega_{;a} = 0$ at initial points.

\subsection{Which computation?}

This paper showed that any computation, including, for example, one
that calculates the digits of $\pi$, corresponds to a class of
spacetimes that obeys the Einstein-Regge equations.   Which
computation corresponds to the world that we see around us?
As noted, quantum cellular automata and random computations are
both reasonable candidates for the `universal' computation.
Cellular automata possess a built-in regular structure which
simplifies the analysis of their behavior.  
A quantum cellular automaton is a natural
choice for a computational substrate, although
care must be taken in the choice of the arrangement
of cells to insure that light propagates at a uniform rate
in all directions.  Bialynicki-Birula has shown that the Weyl,
Dirac, and Maxwell equations can be recovered from quantum
cellular automata on body-centered cubic lattices [62].  

A hyper-diamond lattice consisting of 4-cubes stacked with their
longest diagonal along the `time' direction also represents a natural
lattice in which to embed a CA-like architecture.  (I am endebted to
R. Sorkin for calling this lattice to my attention.)  Each vertex
has four inputs and four outputs.  Label inputs and outputs
from 1 to 4.  The lattice can be filled alternatingly with gates that 
have their two inputs on even lines and their two outputs on odd lines,
and gates that have their two inputs on odd lines and their two outputs 
on even lines.  Lattice gauge theories can be described readily either
by the body-centered cubic lattices of [62] or by hyper-diamond
lattices.

Similarly, a fully
random computational graph with $n$ vertices can be analyzed using 
statistical techniques.  Such a graph possesses a locally tree-like structure;
the smallest loops are of size order $\log n$; and the average loop
has size $\sqrt n$.  Because the smallest loop size gets larger and
larger as $n\rightarrow\infty$, a random computational graph gives
a geometry that is locally approximately flat.  Random graphs tend
to have a complex topology, while CA-like architectures have a
simple topology.

It must be admitted, however, that in investigating quantum cellular
automata and random graphs, we do so exactly because they have
symmetries or statistical regularities that simplify the analysis.
In point of fact, we do not know what form of the `mother computation'
takes.  Rather, we should investigate different candidates and
compare the results of those investigations with observation.   

In fact, computational universality --- the ability of quantum computers 
to simulate each other efficiently --- suggests that it may not
be so important which computation we use.  Computational
universality allows a quantum computer to give rise to all possible 
computable structures.  Consider, for example, a quantum
Turing machine whose input tape is prepared in an equal superposition
of all possible programs.   (In fact, almost any input state for the tape
will give rise to an almost equal superposition of all possible 
programs).  The amplitude for a given input program of length $\ell$
is then just equal to $2^{-\ell/2}$.  Such a quantum computer will produce
all possible outputs in quantum superposition, with an amplitude for
each output equal to the sum of the amplitudes for the programs
that produce that output:
${\cal A}(o) = \sum_{p: U(p) = o} 2^{-\ell_p/2} {\cal A}_p$.   
${\cal A}(o)$ is the {\it algorithmic amplitude} for
the output $o$.  Here, ${\cal A}_p$ is defined as in section (1)
to be the amplitude that the computation starting from program
$p$ actually gives the output $o$.  

To relate the quantum Turing machine picture to the quantum circuit
picture described here, consider a quantum Turing machine
whose outputs are quantum circuits.  The circuits
generated by such a machine constitutes a `family.'  
Some of the circuits in the family could be infinite, even if 
the program producing them is finite in length.  A uniform
superposition of inputs for such a Turing machine produces a 
superposition of quantum circuits, each weighted by their
algorithmic amplitude.  

Because of computational universality, algorithmic 
amplitude is approximately machine
independent: each universal quantum computer can be programmed 
simulate any other quantum computer by a program of finite length.
As a result, the algorithmic amplitude of an output on one universal
quantum computer differs by at most a constant multiplicative factor
from the algorithmic amplitude of the output on another universal
quantum computer (compare the analogous classical result for algorithmic
probability).  Regular circuits such as CAs that produce simple
dynamical laws have high weight in the universal family because
they are produced by short programs.  

Such a universal quantum computer that computes all possible results,
weighted by their algorithmic amplitude, preferentially produces 
simple dynamical laws.  Our own universe apparently obeys simple
dynamical laws, and could plausibly be produced by such a universal
quantum computer.

Quantum computational universality is a powerful feature: it is what
allows the computational universe to reproduce the behavior of any
discrete, local quantum system, including lattice gauge theories.
But care must be taken in applying this power.
In particular, to reproduce a lattice gauge theory, a quantum
computation uses quantum logic gates to reproduce the local
infinitesimal Hamiltonian dynamics.  If those logic gates possess the same
gauge symmetries as the theory to simulated, then the action of the
logic gates can be the same as the action of the simulated theory, and
the dynamics of the simulation can be essentially indistinguishable
from the dynamics of the system simulated.  This is the situation
envisaged by Feynman in his paper on universal quantum simulation [2].
For example, as noted above, the
the simple `swap' model of quantum computation presented in 
section (1) possesses a local $SU(3) \times U(1)$ symmetry at each
logic gate, corresponding to transformations on the triplet
and singlet subspaces, together with a local $SU(2)$ symmetry
on the wires.  Whether this local $SU(3) \times SU(2) \times U(1)$
symmetry can be identified with the 
$SU(3) \times SU(2) \times U(1)$ Yang-Mills gauge symmetry of
the standard model will be investigated elsewhere.

\begin{figure}
\begin{center}
\includegraphics[width=300pt]{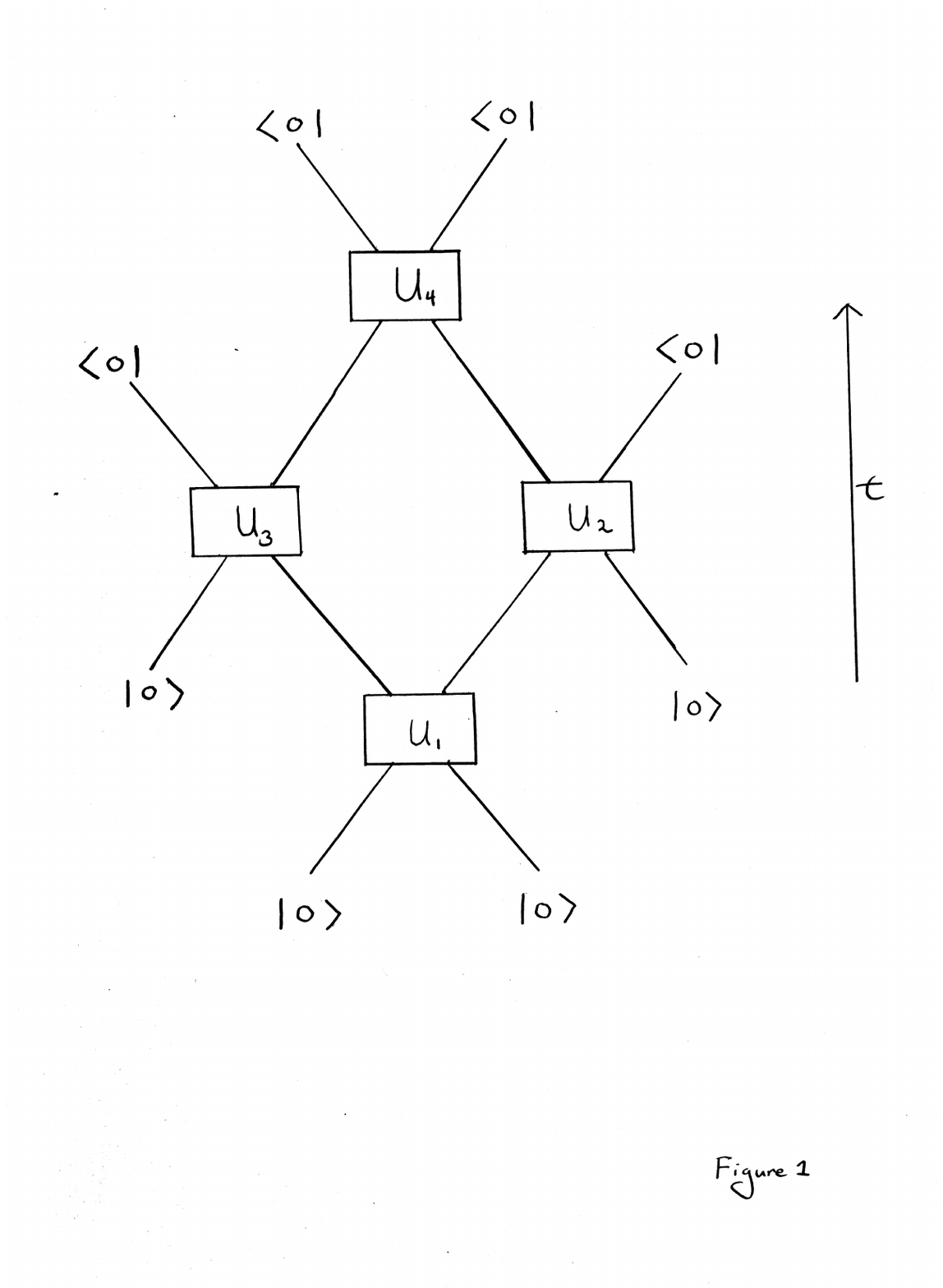}
\caption{
A quantum computation corresponds to a directed, acyclic graph.
Initial vertices correspond to initial states $|0\rangle$;
edges correspond to quantum wires;
internal vertices correspond to quantum logic gates that apply
unitary transformations $U_\ell$;
final vertices correspond to final states $\langle 0|$.}
\label{fig:fig1}
\end{center}
\end{figure}

\begin{figure}
\begin{center}
\includegraphics[width=300pt]{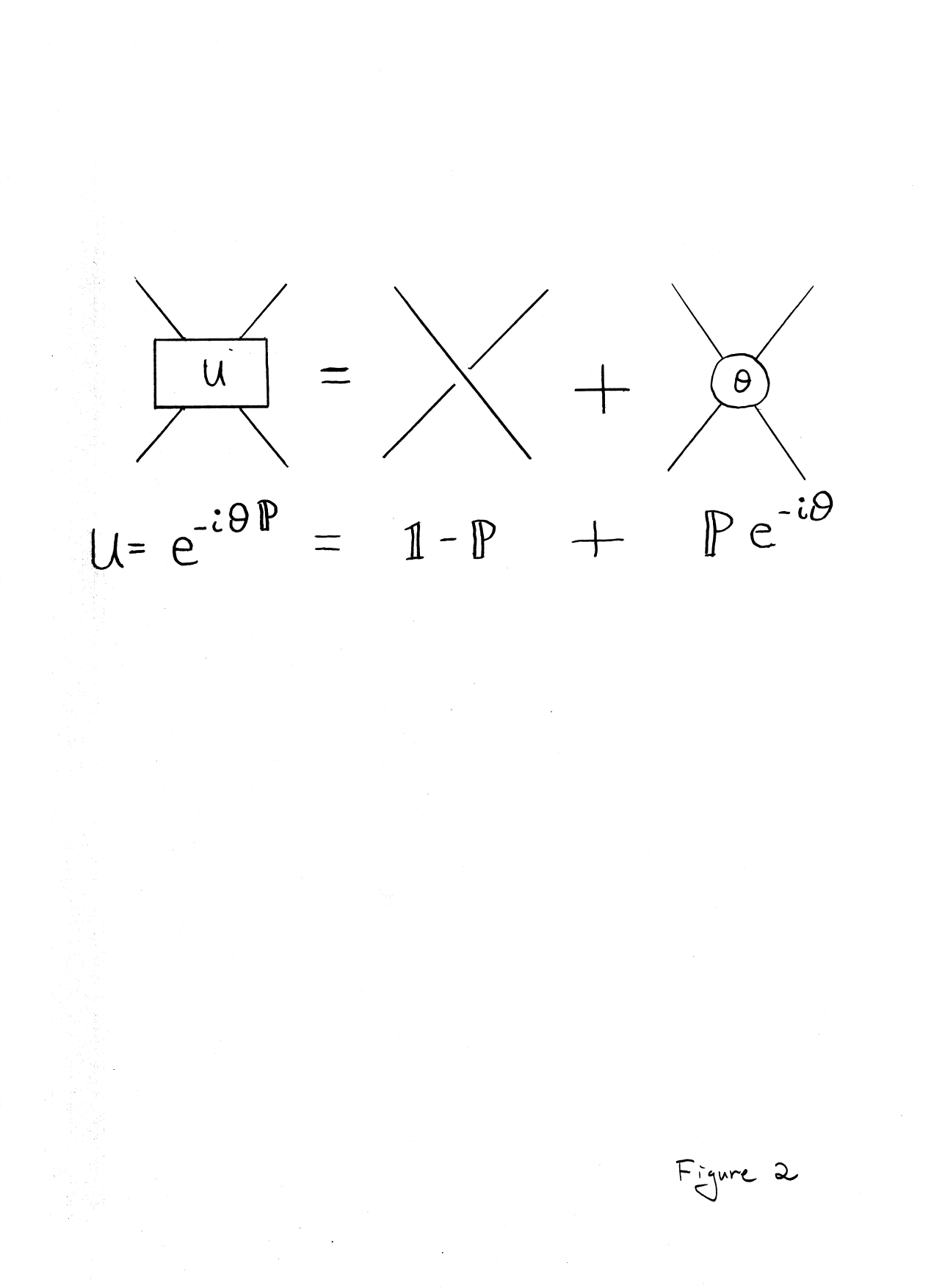}
\caption{
As qubits pass through a quantum logic gate, they can either
scatter or not.  If they scatter, then their state 
acquires a phase $\theta$; if they don't scatter, no phase
is acquired.  Scattering corresponds to an `event'; no
scattering corresponds to a `non-event.'}
\label{fig:fig2}
\end{center}
\end{figure}

\begin{figure}
\begin{center}
\includegraphics[width=300pt]{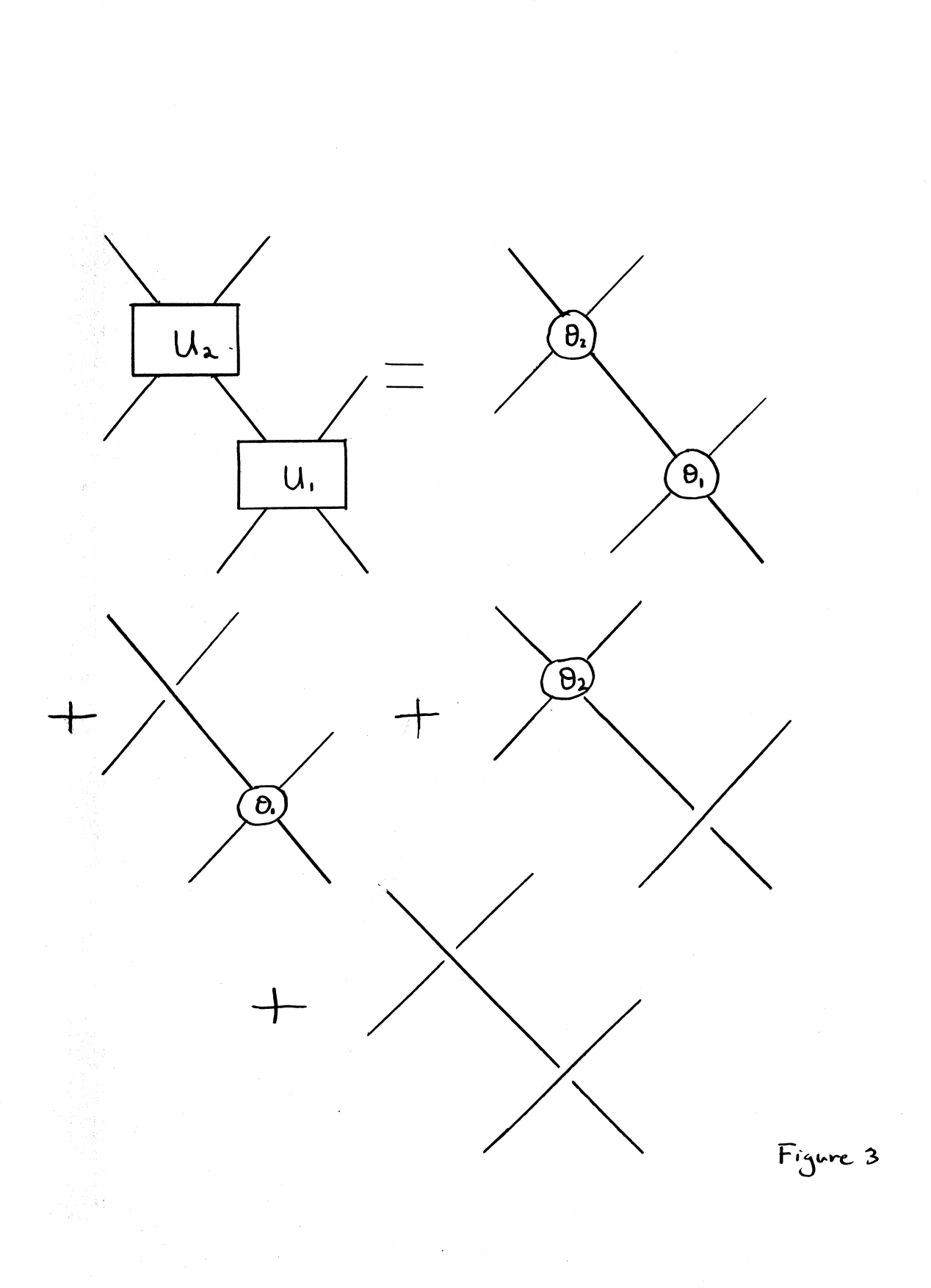}
\caption{
A quantum computation can be decomposed into a superposition
of computational histories, each of which corresponds to a particular pattern
of scattering events.  Each computational history in turn
corresponds to a spacetime with a definite metric.
}
\label{fig:fig3}
\end{center}
\end{figure}

\begin{figure}
\begin{center}
\includegraphics[width=300pt]{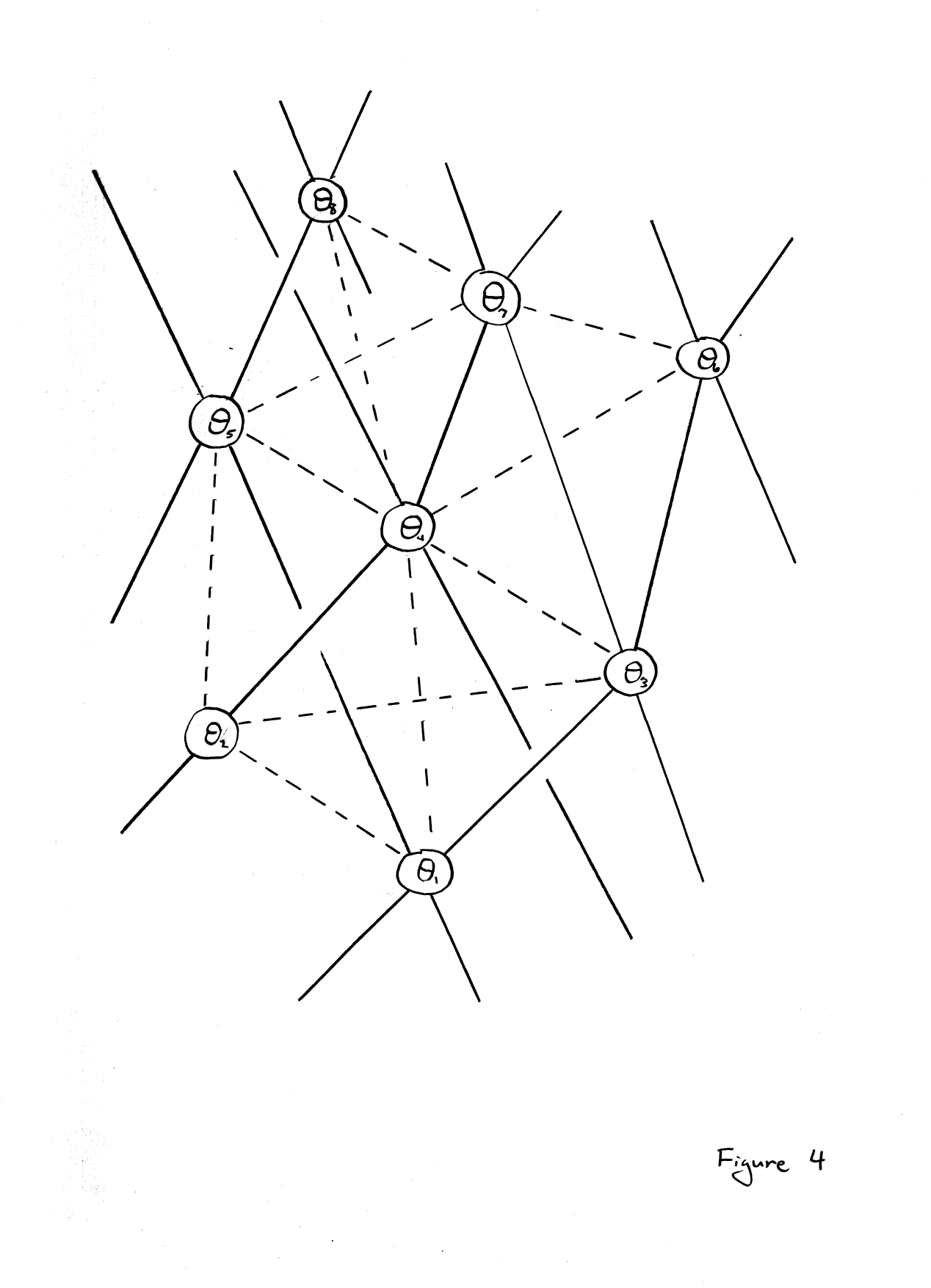}
\caption{
To construct a discretized spacetime from a computational history,
add edges and vertices to form a simplicial `geodesic dome' lattice. 
Figure 4a shows the four-simplex associated with a vertex and
its nearest neighbors.  Figure 4b shows a triangulation of 
a computational history, analogous to the triangulation performed
by a surveyor who adds additional reference vertices and
edges to construct a simplicial lattice.  Edge lengths are defined
by the causal structure of the computational history, 
together with the local action $\hbar\theta_\ell$
of the underlying computation.  The resulting discrete geometry
obeys the Einstein-Regge equations.
}
\label{fig:fig4}
\end{center}
\end{figure}

\end{document}